\documentclass[12pt,a4paper,twoside]{article}
\newenvironment{changemargin}[2]{%
\begin{list}{}{%
\setlength{\leftmargin}{#1}%
\setlength{\rightmargin}{#2}%
}%
\item[]}
{\end{list}}
\usepackage{graphicx}
\usepackage{color}
\usepackage{lscape}
\usepackage{hyperref}
\usepackage{epstopdf}
\usepackage{lscape}
\usepackage{amsmath}
\usepackage{hyperref}
\usepackage{amsmath}
\usepackage{setspace}
\begin{document}
\baselineskip=0.30in
{\bf \LARGE
\begin{changemargin}{-1.2cm}{0.5cm}
\begin{center}
{Eigensolution techniques, their applications and the Fisher's information entropy of Tietz-Wei diatomic molecular model}
\end{center}
\end{changemargin}}

\vspace{4mm}
\begin{center}
{\Large{\bf B. J. Falaye $^a$$^{,}$$^\dag$$^{,}$}}\footnote{\scriptsize E-mail:~ fbjames11@physicist.net.\\ \dag{Corresponding} author Tel.: +2348103950870}\Large{\bf ,} {\Large{\bf K. J. Oyewumi $^b$$^{,}$}}\footnote{\scriptsize E-mail:~ kjoyewumi66@unilorin.edu.ng}\Large{\bf ,} {\Large{\bf S. M. Ikhdair $^c$$^{,}$}}\footnote{\scriptsize E-mail:~ sikhdair@neu.edu.tr;~ sikhdair@gmail.com.} \Large{\bf and} {\Large{\bf M. Hamzavi $^d$$^{,}$}}\footnote{\scriptsize E-mail:~ majid.hamzavi@gmail.com}
\end{center}
{\small
\begin{center}
{\it $^\textbf{a}$Applied Theoretical Physics Division, Department of Physics, Federal University Lafia,  P. M. B. 146, Lafia, Nassarawa State Nigeria.}
{\it $^\textbf{b}$Theoretical Physics Section, Department of Physics, University of Ilorin,  P. M. B. 1515, Ilorin, Nigeria.}
{\it $^\textbf{c}$Department of Physics, Faculty of Science, an-Najah National University, New campus, P. O. Box 7, Nablus, West Bank, Palestine.}
{\it $^\textbf{d}$Department of Physics, University of Zanjan, Zanjan, Iran.}
\end{center}}

\begin{center}
\textbf{Phys. Scr. 89 (2014) 115204 (27pp)}
\end{center}

\begin{abstract}
\noindent
In this study, approximate analytical solution of Schr\"odinger, Klein-Gordon and Dirac equations under the Tietz-Wei (TW) diatomic molecular potential are represented by using an approximation for the centrifugal term. We have applied three types of eigensolution techniques; the functional analysis approach (FAA), supersymmetry quantum mechanics (SUSYQM) and asymptotic iteration method (AIM) to solve Klein-Gordon, Dirac and Schr\"odinger equations, respectively. The energy eigenvalues and the corresponding eigenfunctions for these three wave equations are obtained and some numerical results and figures are reported. It has been shown that these techniques yielded exactly same results.  some expectation values of the TW diatomic molecular potential within the framework of the Hellmann-Feynman theorem (HFT) have been presented. The probability distributions which characterize the quantum-mechanical states of TW diatomic molecular potential are analyzed by means of complementary information measures of a probability distribution called the Fishers information entropy. This distribution has been described in terms of Jacobi polynomials, whose characteristics are controlled by the quantum numbers.
\end{abstract}


eigensolution techniques; Fisher's information entropy; 

{\bf PACs}: 03.65.–w, 03.65.Ge, 03.67.–a; 03.70.+k; 03.65.-a.

\noindent
\begin{changemargin}{-1.2cm}{0.7cm}
\section{Introduction}
During the past decades, solutions of wave equations in relativistic and non-relativistic quantum mechanics have received great attention. Whether in the relativistic or non-relativistic regimes, the exact solutions are only possible for some special cases of interactions due to the inverse square centrifugal term appearing in the realistic three-dimensional space. Therefore, for many potential models one has to use some approximations for the centrifugal term. Among these approximations, the frequently used Pekeris approximation \cite{Y1, Y2} introduces an exponential approximation to the centrifugal barrier. The TW diatomic molecular potential is one of those potentials which cannot be exactly solved. This potential was proposed as an intermolecular potential and is one of the best potential models considered to describe the vibrational energy of diatomic molecules \cite{Y3, Y4}. 

In one of the interesting works of Gordillo-Vizquez and Kunc \cite{T1}, they derived radial probability distributions of diatomic molecules in excited rotational-vibrational states using the Bohr-Sommerfeld quantization rule and the Hamilton-Jacoby theory. They compared the obtained semiclassical distributions for the rotating Morse and TW oscillators, in a broad range of rotational and vibrational quantum numbers, with one another and with quantum-mechanical distributions obtained from numerical solution of the Schr\"{o}dinger equation using a Rydberg-Klein-Rees (RKR) potential. In the recent years, Hamzavi et al \cite{T2} have presented energy eigenvalues and corresponding eigenfunctions in closed form. The authors present some remarks and numerical results for some diatomic molecules. Shortly thereafter, the bound state solutions of the relativistic Klein-Gordon equation with the TW diatomic molecular potential have been presented for the s wave by Sun and Dong \cite{T3}. It is shown that the solutions can be expressed by the generalized hypergeometric functions.

Apart from the works done on TW diatomic molecular potential, many authors have solved the Schr\"{o}dinger equation in various potential fields \cite{T4}-\cite{GH2}. Despite all several worthy attempt in solving Schr\"{o}dinger equation with several model potentials, the basic information-theoretic quantities remain to be computed. This is because of the lack of knowledge in the information-theoretic properties of special functions. In this paper, our aim is to solve the nonrelativistic Schr\"{o}dinger equations under the TW diatomic molecular potential by via the AIM, by considering the Pekeris approximation. We have also calculated some expectation values corresponding to the TW diatomic molecular potential  using the Hellmann-Feynman theorem. 

We proceed further to study the TW diatomic molecular potential within the framework of relativistic Klein-Gordon and Dirac equation. It is worth to be noted that the Klein-Gordon equation correctly describes the spinless {pion}\footnote{The pion is a composite particle; no spinless elementary particles have yet been found, although the Higgs boson is theorized to exist as a spin-zero boson, according to the Standard Model.}. The Dirac equation provided a description of elementary spin-1/2 particles, such as electrons, consistent with both the principles of quantum mechanics and the theory of special relativity, and made relativistic corrections to quantum mechanics \cite{T16,T17}. These relativistic equations have received a lot of interest from several authors in both Theoretical Physics, Nuclear Physics and related areas. In this study, these relativistic equations with TW diatomic molecular potential shall be studied extensively.

Furthermore, the probability distributions are analyzed by means of complementary information measures of a probability distribution called as the Fishers information entropy. Given a normalized-to-unity (probability) density $\rho$ on $\Re$, we can define several functions of $\rho$ measuring the uncertainty or information content associated with this
density. Thus, Fisher's information is a derivative functional of the density, so that it is very sensitive to local rearrangements of $\rho$. Applications to a large variety of problems in theoretical physics have received a strong impulse when it was realized that the spatial distribution of the single-particle probability density $\rho(\vec{r})$ of a many-particle system, which is the basic variable of the density functional theory \cite{K1,Y5,Y6,Y7} can be quantitatively measured by its translationally invariant Fisher information in a different and complementary manner as the Shannon entropy. Both quantities characterize the information theoretic content of the density $\rho(\vec{r})$ which describes the physical state under consideration \cite{Y7}. The Fisher information has been used in present study to analyze probability distribution of the approximate solution obtained for the TW diatomic molecular potential.

The organization of this paper is as follows: in the next section we briefly introduce FAA, SUSYQM and AIM. In section \ref{sec3}, the non-relativistic solutions of Schr\"odinger equation with the TW diatomic molecular potential are presented and some expectation values are reported. We have obtained the energy eigenvalues and the wave functions of the Klein-Gordon and Dirac equations in section \ref{sec4}. Information-theoretic measures for TW diatomic molecular potential is given in section  \ref{sec5}. Section \ref{sec6} present the numerical discussions and finally, our conclusion is given in section  \ref{sec3}.

\section{Eigensolution techniques}
In quantum mechanics, while solving relativistic and nonrelativistic wave equation in the presence of some typical central or non central potential model, we do often come across differential equation of the form
\begin{equation}
\Psi''(s)+\frac{(k_1-k_2s)}{s(1-k_3s)}\Psi'(s)+\frac{(As^2+Bs+C)}{s^2(1-k_3s)^2}\Psi(s)=0,
\label{E1}
\end{equation}
after which an appropriate coordinate transformation of the form $s=s(r)$ has been used. In other to find energy eigenvalues and the wavefunction (eigensolution) of the second-order homogeneous linear differential equations of the above form, there have been several eigensolution techniques developed to exactly solve the quantum systems. Some of this techniques include the AIM \cite{BJ1}-\cite{N21}, Feynman integral formalism \cite{BJ3,BJ4,BJ5}, FAA\cite{BJ6, BJ7, BJ8, BJ9, BJ10, BJ11}, exact quantization rule method \cite{BJ12, BJ13, BJ14, BJ15, BJ16, BJ17}, proper quantization rule \cite{BJ18, BJ20}, Nikiforov-Uvarov (NU) method \cite{S1,H1,BJ21, BJ22, BJ23, BJ24, BJ25, BJ26, BJ27, BJ28,BJ29}, SUSYQM \cite{BJ30,BJ31,BJ32,BJ33,BJ34,BJ35,BJ36,BJ37,BJ38,BJ39,BJ40,BJ41,BJ42,BJ42,BJ43,BJ44,BJ45, ES1}, the ansatz method \cite{H3, H4, H5, H6}, etc. 

The beauty about these techniques is that when they are apply to solve equation of the form (\ref{E1}), they yield exactly the same eigensolution results, despite the fact that the calculation approaches varies from one technique to the other. In this work we focus our attention unto three of these techniques: FAA, SUSYQM and AIM. We apply AIM to find the eigensolution of nonrelativistic Schr\"{o}dinger equation. Furthermore, the eigensolutions of the relativistic Klein-Gordon and Dirac equations are obtain via FAA and SUSYQM respectively. We have demonstrated that the three techniques yield exactly the same results.

FAA also called the traditional method by some authors has been introduced for solving equation of the form (\ref{E1}) ever since the birth of quantum mechanics. In the approach, one transform the equation of form (\ref{E1}) to a form of hypergeometric equation $_2F_1(a, b; c; s)$ via an appropriate transformation by considering the singularity points of the differential equation within the framework of the Frebenius theorem. The eigensolutions are obtained from the properties of the hypergeometric functions on the condition that the series $_2F_1(a, b; c; s)$ approaches infinity unless $a$ is a negative integer (i.e $a=-n$). This method have been employed by alot of researchers to obtain eigensolution of quantum mechanical problems in both relativistic and non relativistic case \cite{BJ6, BJ7, BJ8, BJ9, BJ10, BJ11}

For some decades ago, the ideas of SUSYQM have been applied to alot of quantum mechanical problems (both in the relativistic and nonrelativistic case). In 1982, path integral formulation of SUSYQM was first given by Salomonson and van Holten \cite{BJ34}. There after, by using SUSY methods, some authors reveal that the tunneling rate through double well barriers could be accurately determined \cite{BJ46,BJ48,BJ49,BJ50}. In the recent years, there has been alot of efforts to extend the ideas of SUSYQM to higher dimensional systems as well as to systems with large numbers of particles with a motivation of understanding the potential problems of widespread interest in nuclear, atomic, statistical and condensed
matter physics \cite{BJ30, BJ31,BJ32,BJ33,BJ34,BJ35,BJ36,BJ37,BJ38,BJ39,BJ40,BJ41,BJ42,BJ43,BJ44,BJ45}. The concept of shape invariant potential was introduced within the framework of SUSYQM in 1983 by  Gendenshtein \cite{BJ42}. Gendenshtein shows that for any shape invariant potential, the energy eigenvalue spectra could be obtained algebraically. Very recently,  the bound-state spectra for some potentials with unbroken and broken supersymmetry are investigated by the quantization condition of AIM \cite{RR1}.

The asymptotic iteration method (AIM) to find eigensolution of equation of the form (\ref{E1}) was introduced by Ciftci et al. \cite{N1,N2} in 2003 and 2005. Ever since then, it has been used in many physical systems to obtain the whole spectra [29-50]. The beauty about the method is that it reproduces exact solutions of
many exactly solvable quantum systems and also gives accurate result for the non solvable potentials such as sextic oscillator, deformed Coulomb potential, etc.
 
In the subsequent sections, we apply these three techniques to solve various quantum mechanical problems and shows that they yield exactly same results.

\section{Nonrelativistic Solutions of Schr\"{o}dinger equation with the TW diatomic molecular potential} \label{sec3}
It is well known that exact solutions play an important role in quantum mechanics since they contain all the necessary information regarding the quantum system under consideration. In this section we obtain the bound states solution of the Schr\"{o}dinger equation for the TW diatomic molecular potential. For this purpose we write the Schr\"{o}dinger equation 
in three dimension as
\begin{equation}
\left\{-\frac{\hbar^2}{2\mu}\left[\frac{1}{r^2}\frac{\partial}{\partial r}r^2\frac{\partial}{\partial r}+\frac{1}{r^2\sin \theta}\frac{\partial}{\partial\theta}\left(\sin\theta\frac{\partial}{\partial\theta}\right)+\frac{1}{r^2\sin^2\theta}\frac{\partial^2}{\partial\phi^2}\right]+V(r)-E\right\}\psi(r,\theta,\phi)=0,
\label{E2}
\end{equation}
where $n$ and $\ell$ denote the radial and orbital angular momentum quantum numbers, $r$ is the internuclear separation of the diatomic molecules, and $E_{n\ell}$ is the bound-state energy eigenvalues. The $\mu$ and $V(r)$ represent the reduced mass and interaction potential respectively. By taking $\psi(r,\theta,\phi)=r^{-1}U_{n\ell}(r)Y_{\ell m}(\theta,\phi)$, we can separate the above Schr\"{o}dinger equation via the variable separable technique. Thus, we obtain the following radial Schr\"{o}dinger equation:
\begin{equation}
\frac{d^2U_{n\ell}(r)}{dr^2}+\frac{2\mu}{\hbar^2}\left(E_{n\ell}-V(r)\right)U_{n\ell}(r)=0.
\label{E3}
\end{equation}
\subsection{Any $\ell-$state solution of the radial Schr\"{o}dinger equation with the TW diatomic molecular potential: Asymptotic iteration method}
The TW diatomic molecular potential we examine in this study is defined as \cite{T1,T2,T3,POT2}
\begin{equation}
V(r)=D\left[\frac{1-e^{-b_h(r-r_e)}}{1-c_he^{-b_h(r-r_e)}}\right]^2,
\label{E4}
\end{equation}
\begin{figure}[!htb]
\centering \includegraphics[height=100mm,width=180mm]{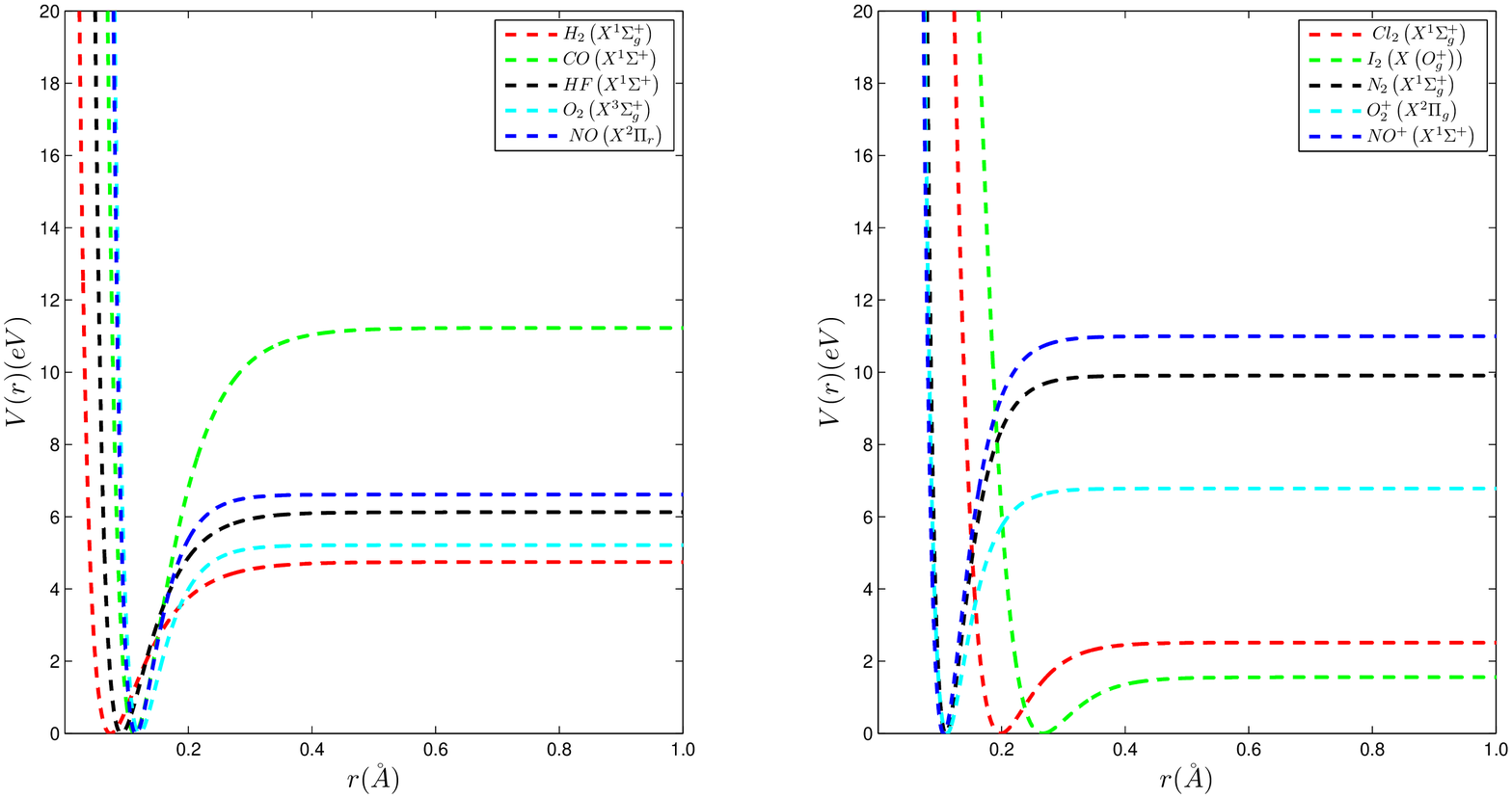}
\caption{{\protect\footnotesize Shape of TW diatomic molecular potential for different diatomic molecules.}}
\label{fig1}
\end{figure}
\begin{table}[!t]
{\scriptsize
\caption{ \small Bound-state energy eigenvalues for $H_2 \left(X^1\Sigma^+_g\right)$, $CO \left(X^1\Sigma^+\right)$, $HF \left(X^1\Sigma^+\right)$, $O_2 \left(X^3 \Sigma^+_g\right)$ and $NO \left(X^2 \Pi_r\right)$ molecules for various $n$ and rotational $\ell$ quantum numbers in TW diatomic molecular potential. } \vspace*{10pt}{
\begin{tabular}{ccccccc}\hline\hline
{}&{}&{}&{}&{}&{}&{}\\[-1.0ex]
$n$&$\ell$&$H_2 \left(X^1\Sigma^+_g\right)$&$CO \left(X^1\Sigma^+\right)$&$HF \left(X^1\Sigma^+\right)$&$O_2 \left(X^3 \Sigma^+_g\right)$&$NO \left(X^2 \Pi_r\right)$\\[2.5ex]\hline\hline
0	&0	&0.269255177	&0.150703328	&0.253867064	&0.097520341	&0.117508622\\[1ex]
1	&0	&0.778774368	&0.448390501	&0.742676288	&0.289698792	&0.349308434\\[1ex]
	&1	&0.792535216	&0.448858605	&0.747534657	&0.290048441	&0.349720567\\[1ex]
2	&0	&1.252418022	&0.741348914	&1.207523735	&0.478099389	&0.576841245\\[1ex]
	&1	&1.265398618	&0.741811764	&1.212186887	&0.478444568	&0.577248398\\[1ex]
	&2	&1.291246200	&0.742737445	&1.221507324	&0.479134962	&0.578062645\\[1ex]
3	&0	&1.691333219	&1.029610568	&1.648826110	&0.662728172	&0.800110112\\[1ex]
	&1	&1.703567986	&1.030068193	&1.653297607	&0.663068932	&0.800512223\\[1ex]
	&2	&1.727931688	&1.030983431	&1.662234886	&0.663750364	&0.801316967\\[1ex]
	&3	&1.764215767	&1.032356253	&1.675626521	&0.664772574	&0.802523312\\[1ex]
4	&0	&2.096613542	&1.313207150	&2.066989237	&0.843591024	&1.019118256\\[1ex]
	&1	&2.108134592	&1.313659582	&2.071272547	&0.843927335	&1.019515545\\[1ex]
	&2	&2.131078206	&1.314564429	&2.079833592	&0.844599847	&1.020309967\\[1ex]
	&3	&2.165250313	&1.315921667	&2.092661234	&0.845608699	&1.021501665\\[1ex]
	&4	&2.210366885	&1.317731254	&2.109738804	&0.846953691	&1.023090734\\[1ex]
5	&0	&2.469302053	&1.592170028	&2.462408428	&1.020693823	&1.233868623\\[1ex]
	&1	&2.480139341	&1.592617287	&2.466506913	&1.021025634	&1.234261023\\[1ex]
	&2	&2.501722377	&1.593511812	&2.474698450	&1.021689355	&1.235045656\\[1ex]
	&3	&2.533870783	&1.594853558	&2.486972183	&1.022684786	&1.236222354\\[1ex]
	&4	&2.576320598	&1.596642482	&2.503311852	&1.024012093	&1.237791733\\[1ex]
	&5	&2.628731798	&1.598878537	&2.523695846	&1.025671003	&1.239753045\\[1ex]
6	&0	&2.810394062	&1.866530245	&2.835468783	&1.194042501	&1.444364456\\[1ex]
	&1	&2.820575527	&1.866972369	&2.839385715	&1.194369905	&1.444751656\\[1ex]
	&2	&2.840853493	&1.867856625	&2.847214278	&1.195024707	&1.445526678\\[1ex]
	&3	&2.871060545	&1.869182965	&2.858943887	&1.196006859	&1.446688856\\[1ex]
	&4	&2.910951716	&1.870951346	&2.874558705	&1.197316355	&1.448238358\\[1ex]
	&5	&2.960211497	&1.873161728	&2.894037651	&1.198953074	&1.450175034\\[1ex]
	&6	&3.018462601	&1.875814019	&2.917354461	&1.200917082	&1.452498823\\[1ex]
7	&0	&3.120839735	&2.136318554	&3.186545521	&1.363642945	&1.650608943\\[1ex]
	&1	&3.130391433	&2.136755573	&3.190284072	&1.363965886	&1.650991545\\[1ex]
	&2	&3.149416096	&2.137629616	&3.197756011	&1.364611822	&1.651756445\\[1ex]
	&3	&3.177758594	&2.138940639	&3.208951016	&1.365580727	&1.652903645\\[1ex]
	&4	&3.215191967	&2.140688596	&3.223853651	&1.366872458	&1.654433256\\[1ex]
	&5	&3.261423932	&2.142873447	&3.242443361	&1.368487069	&1.656345856\\[1ex]
	&6	&3.316105073	&2.145495102	&3.264694547	&1.370424434	&1.658639934\\[1ex]
	&7	&3.378838214	&2.148553507	&3.290576601	&1.372684523	&1.661316623\\[1ex]
\hline\hline
\end{tabular}\label{Tab3}}
\vspace*{-1pt}}
\end{table}
\begin{table}[!t]
{\scriptsize
\caption{\small Bound-state energy eigenvalues for $Cl_2 \left(X^1\Sigma^+_g\right)$, $I_2 \left(X\left(O_g^+\right)\right)$, $N_2 \left(X^1\Sigma^+_g\right)$, $O_2 ^+\left(X^2 \Pi_g\right)$ and $NO^+ \left(X^1 \Sigma^+\right)$ molecules for various $n$ and rotational $\ell$ quantum numbers in TW diatomic molecular potential. } \vspace*{10pt}{
\begin{tabular}{ccccccc}\hline\hline
{}&{}&{}&{}&{}&{}&{}\\[-1.0ex]
$n$&$\ell$&$Cl_2 \left(X^1\Sigma^+_g\right)$&$I_2 \left(X\left(O_g^+\right)\right)$&$N_2 \left(X^1\Sigma^+_g\right)$&$O_2 ^+\left(X^2 \Pi_g\right)$	&$NO^+ \left(X^1 \Sigma^+\right)$\\[2.5ex]\hline\hline
0	&0	&-0.034819832	&-0.013326639	&-0.146754672	&-0.118620934	&-0.147932212\\[1ex]
1	&0	&-0.105090197	&-0.040121814	&-0.443362676	&-0.358867912	&-0.446647624\\[1ex]
	&1	&-0.105029462	&-0.040112565	&-0.442863155	&-0.358444223	&-0.446145438\\[1ex]
2	&0	&-0.176229866	&-0.067116219	&-0.744144013	&-0.603144924	&-0.749198864\\[1ex]
	&1	&-0.176168730	&-0.067106939	&-0.743639828	&-0.602716512	&-0.748692050\\[1ex]
	&2	&-0.176046445	&-0.067088376 &-0.742631522	&-0.601859914	&-0.747678995\\[1ex]
3	&0	&-0.248235753	&-0.094309263	&-1.049092718	&-0.851448145	&-1.055581449\\[1ex]
	&1	&-0.248174213	&-0.094299951	&-1.048583979	&-0.851015224	&-1.055070323\\[1ex]
	&2	&-0.248051126	&-0.094281326	&-1.047566427	&-0.850149446	&-1.054048112\\[1ex]
	&3	&-0.247866497	&-0.094253391	&-1.046040089	&-0.848850667	&-1.052515317\\[1ex]
4	&0	&-0.321104777	&-0.121700354	&-1.358203204	&-1.103773145	&-1.365790735\\[1ex]
	&1	&-0.321042835	&-0.121691011	&-1.357689794	&-1.103335534	&-1.365275425\\[1ex]
	&2	&-0.320918943	&-0.121672325	&-1.356663018	&-1.102460745	&-1.364244756\\[1ex]
	&3	&-0.320733113	&-0.121644296	&-1.355122889	&-1.101148245	&-1.362698698\\[1ex]
	&4	&-0.320485340	&-0.121606922	&-1.353069424	&-1.099398515	&-1.360637524\\[1ex]
5	&0	&-0.394833886	&-0.149288907	&-1.671469715	&-1.360116144	&-1.679822917\\[1ex]
	&1	&-0.394771542	&-0.149279532	&-1.670951742	&-1.359674023	&-1.679303323\\[1ex]
	&2	&-0.394646854	&-0.149260783	&-1.669915721	&-1.358789934	&-1.678263938\\[1ex]
	&3	&-0.394459821	&-0.149232663	&-1.668361692	&-1.357463945	&-1.676705078\\[1ex]
	&4	&-0.394210451	&-0.149195166	&-1.666289908	&-1.355695956	&-1.674626437\\[1ex]
	&5	&-0.393898743	&-0.149148297	&-1.663700136	&-1.353485989	&-1.672028338\\[1ex]
6	&0	&-0.469420044	&-0.177074336	&-1.988886468	&-1.620473134	&-1.997673567\\[1ex]
	&1	&-0.469357301	&-0.177064931	&-1.988363928	&-1.620026744	&-1.997149548\\[1ex]
	&2	&-0.469231814	&-0.177046120	&-1.987318690	&-1.619133434	&-1.996101546\\[1ex]
	&3	&-0.469043594	&-0.177017907	&-1.985750985	&-1.617793611	&-1.994529545\\[1ex]
	&4	&-0.468792626	&-0.176980288	&-1.983660639	&-1.616007412	&-1.992433758\\[1ex]
	&5	&-0.468478925	&-0.176933265	&-1.981047963	&-1.613774723	&-1.989814254\\[1ex]
	&6	&-0.468102492	&-0.176876837	&-1.977912901	&-1.611095634	&-1.986670600\\[1ex]
7	&0	&-0.544860232	&-0.205056063	&-2.310447874	&-1.884840245	&-2.319338467\\[1ex]
	&1	&-0.544797097	&-0.205046628	&-2.309920665	&-1.884389145	&-2.318809929\\[1ex]
	&2	&-0.544670819	&-0.205027756	&-2.308866326	&-1.883486923	&-2.317753276\\[1ex]
	&3	&-0.544481405	&-0.204999450	&-2.307284746	&-1.882133323	&-2.316168345\\[1ex]
	&4	&-0.544228850	&-0.204961709	&-2.305176089	&-1.880328918	&-2.314055274\\[1ex]
	&5	&-0.543913166	&-0.204914532	&-2.302540413	&-1.878073478	&-2.311414009\\[1ex]
	&6	&-0.543534355	&-0.204857921	&-2.299377826	&-1.87536692	&-2.308244687\\[1ex]
	&7	&-0.543092414	&-0.204791875	&-2.295688306	&-1.872209698	&-2.304547249\\[1ex]
\hline\hline
\end{tabular}\label{Tab4}}
\vspace*{-1pt}}
\end{table}
\begin{figure}[!t]
\centering \includegraphics[height=100mm,width=180mm]{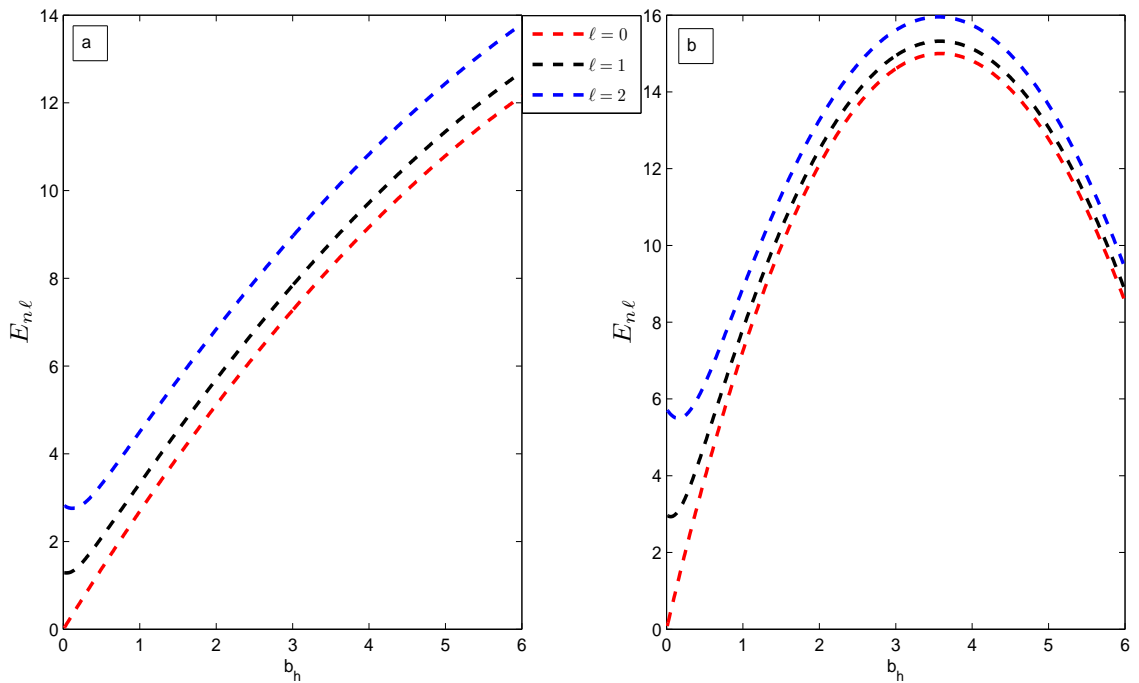}
\caption{{\protect\small (a) The variation of the ground state energy spectrum for various values of $\ell$ as a function of the parameter $b_h$. We choose $c_h=0.03$, $\mu=1$,  $r_e=1.207$ and $D=15$. (b) The variation of the first excited energy state for various $\ell$ as a function of the parameter $b_h$}}
\label{fig2}
\end{figure}
\begin{figure}[!t]
 \includegraphics[height=100mm,width=180mm]{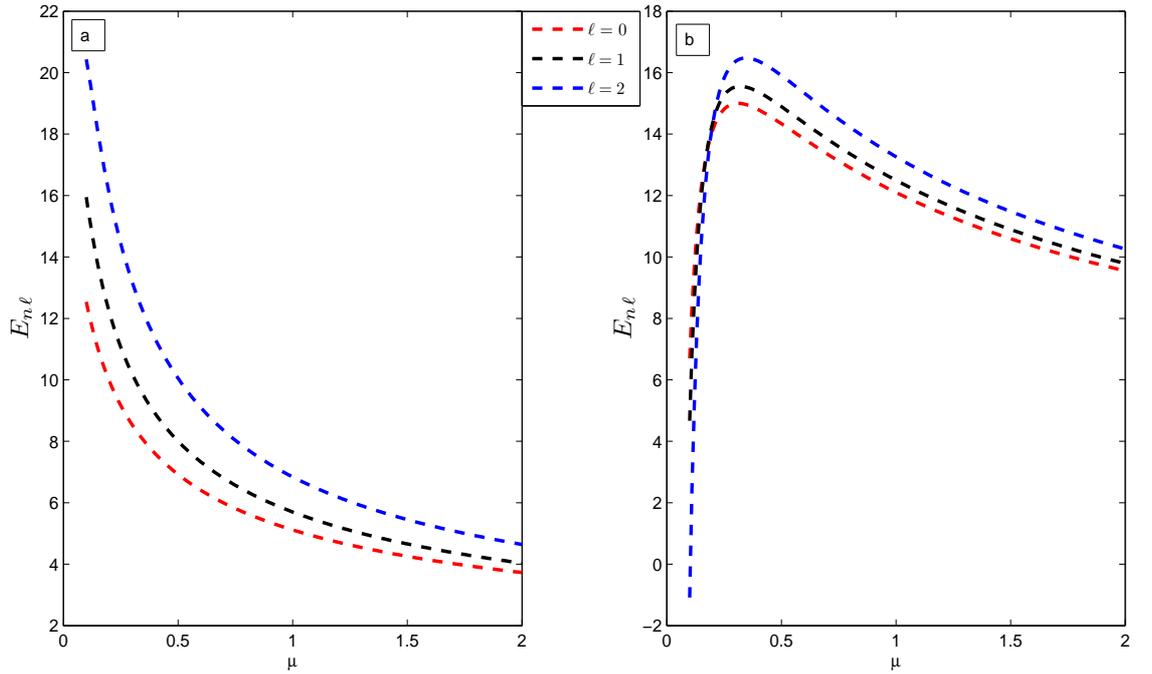}
\caption{{\protect\small (a) The variation of the ground state energy spectrum for various values of $\ell$ as a function of the particle mass $\mu$. We choose $c_h=0.03$, $b_h=2$,  $r_e=1.207$ and $D=15$. (b) The variation of the first excited energy state for various $\ell$ as a function of the particle mass $\mu$}}
\label{fig3}
\end{figure}
\begin{figure}[!t]
 \includegraphics[height=100mm,width=180mm]{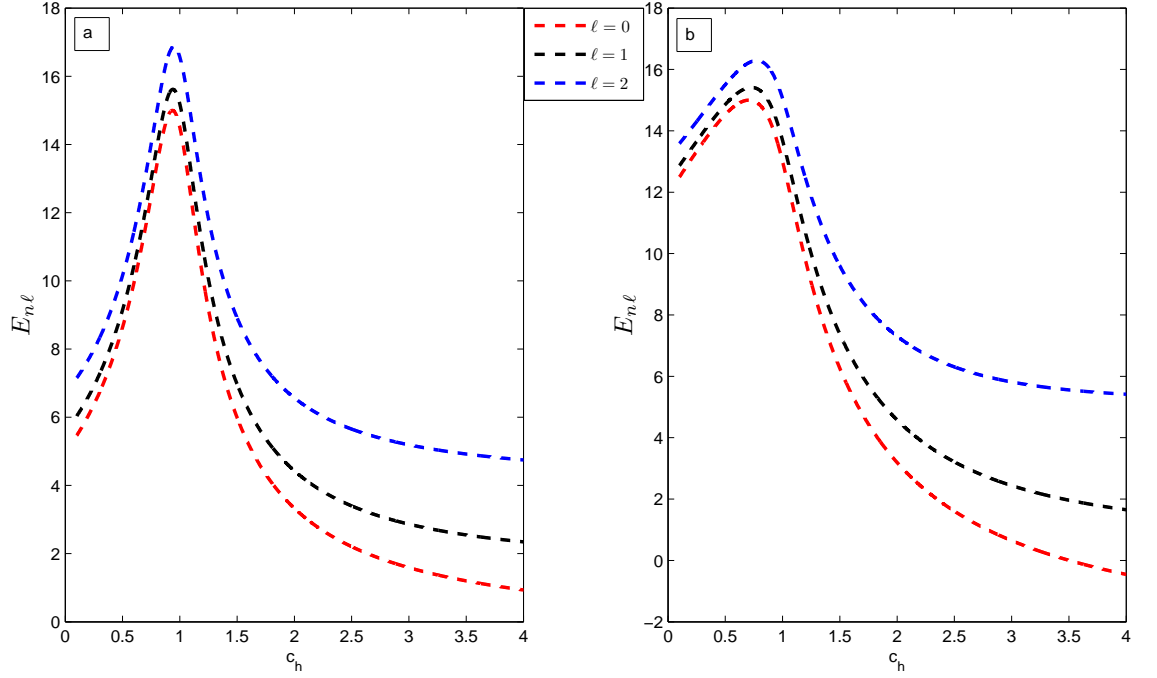}
\caption{{\protect\small (a) The variation of the ground state energy spectrum for various values of $\ell$ as a function of the potential constant $c_h$. We choose $\mu=1$, $b_h=2$,  $r_e=1.207$ and $D=15$. (b) The variation of the first excited energy state for various $\ell$ as a function of the potential constant $c_h$}}
\label{fig4}
\end{figure}
\begin{figure}[!t]
\includegraphics[height=100mm,width=180mm]{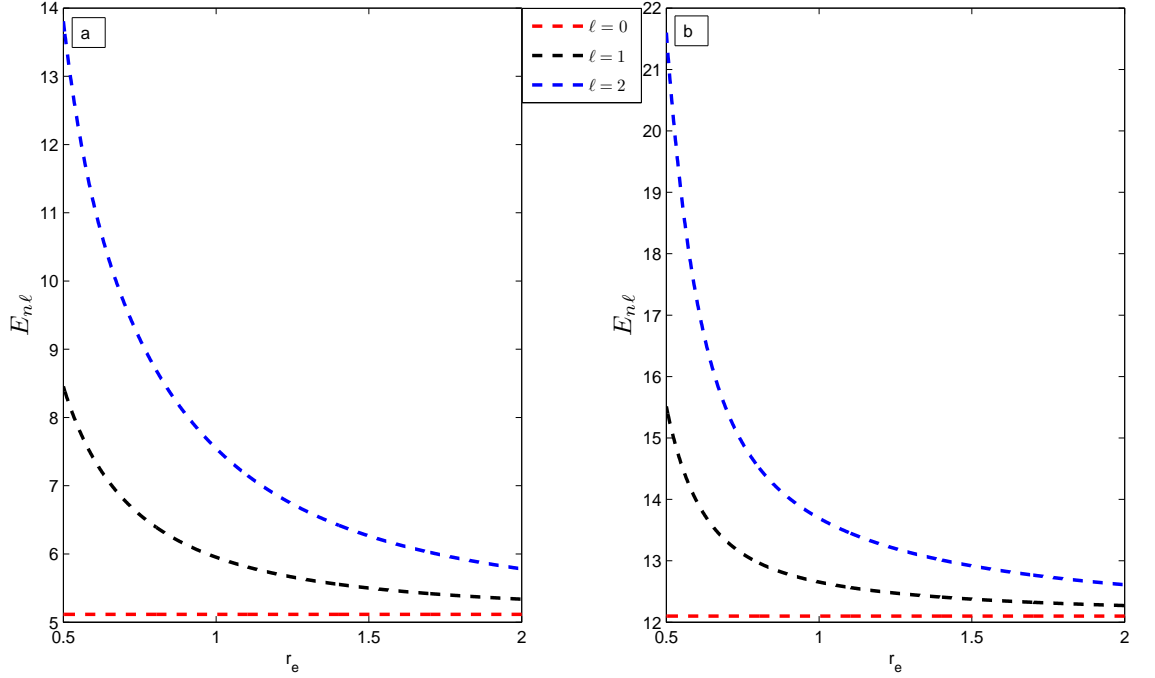}
\caption{{\protect\small (a) The variation of the ground state energy spectrum for various values of $\ell$ as a function of the molecular bond length $r_e$. We choose $\mu=1$, $b_h=2$,  $c_h=0.03$ and $D=15$. (b) The variation of the first excited energy state for various $\ell$ as a function of the molecular bond length $r_e$}}
\label{fig5}
\end{figure}
\begin{figure}[!t]
\includegraphics[height=100mm,width=180mm]{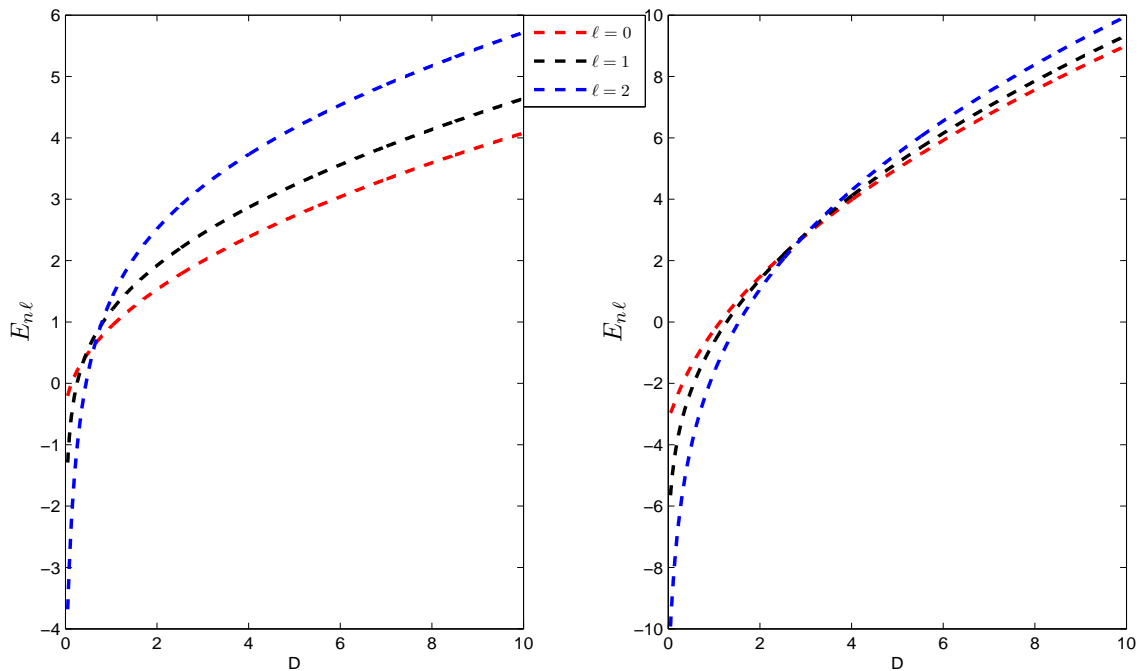}
\caption{{\protect\small (a) The variation of the ground state energy spectrum for various values of $\ell$ as a function of the potential well depth $D$. We choose $\mu=1$, $b_h=2$,  $c_h=0.03$ and $r_e=1.207$. (b) The variation of the first excited energy state for various $\ell$ as a function of the potential well depth $D$}}
\label{fig6}
\end{figure}
\begin{figure}[!t]
\includegraphics[height=100mm,width=180mm]{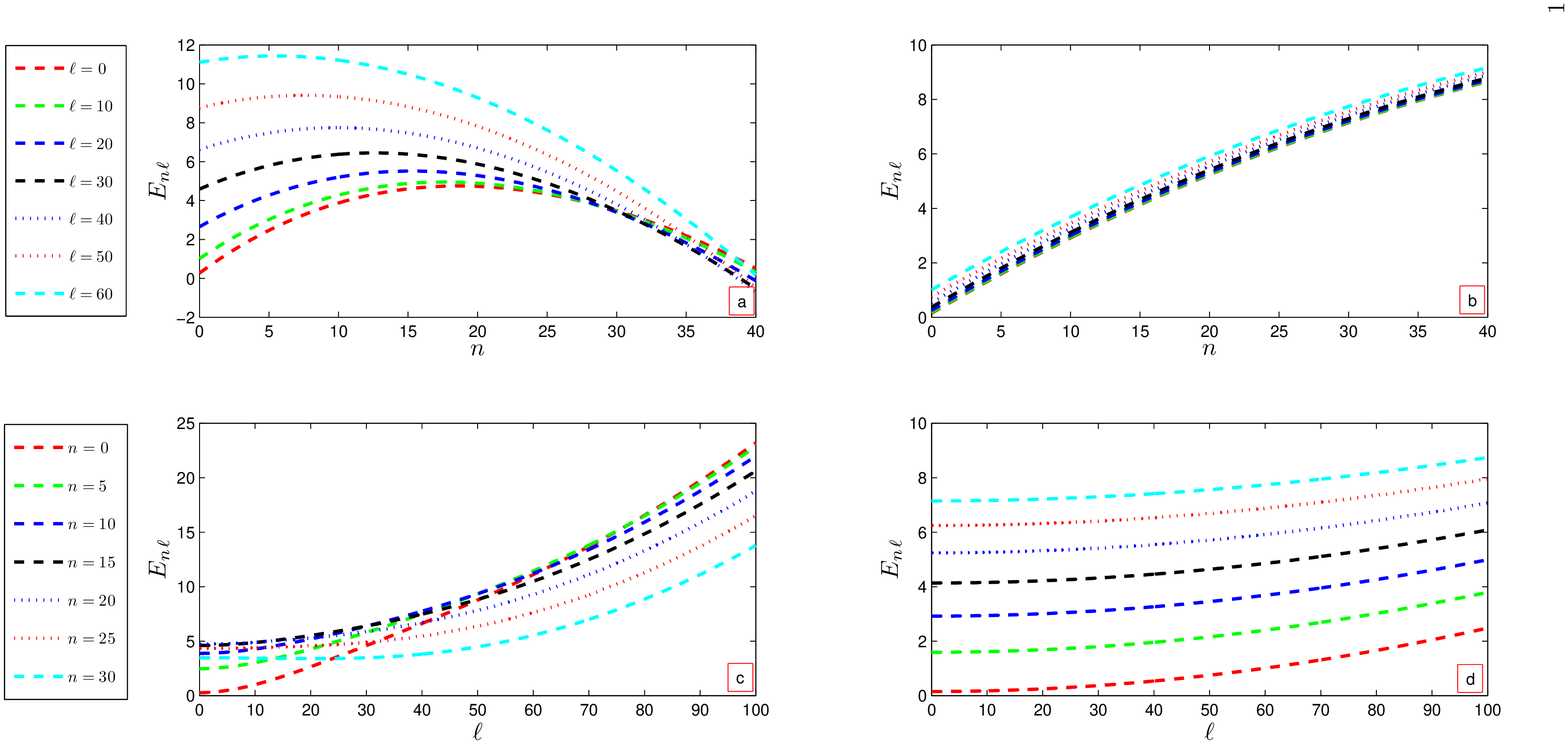}
\caption{{\protect\small (a) The variations of energy (in eV) in TW diatomic molecular potential, with respect to vibrational ($n$) for $H_2 \left(X^1\Sigma^+_g\right)$ molecules. (b) Same as (a) but for $CO \left(X^1\Sigma^+\right)$ (c) The variations of energy (in eV) in TW diatomic molecular potential, with respect to rotational ($\ell$) for $H_2 \left(X^1\Sigma^+_g\right)$ molecules. (d) Same as (c) but for $CO \left(X^1\Sigma^+\right)$.}}
\label{fig7}
\end{figure}
\begin{figure}[!t]
\includegraphics[height=100mm,width=180mm]{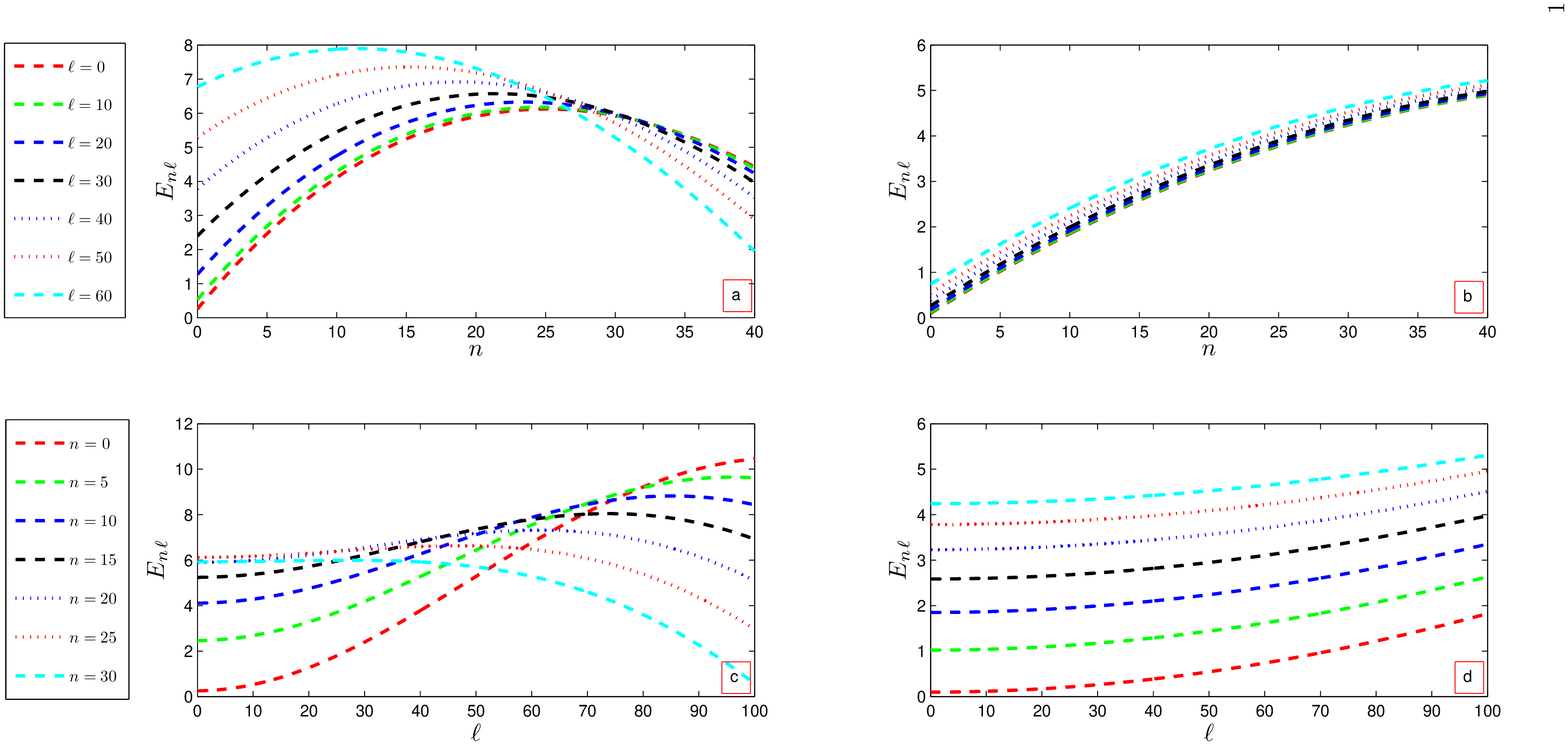}
\caption{{\protect\small (a) The variations of energy (in eV) in TW diatomic molecular potential, with respect to vibrational ($n$) for $HF \left(X^1\Sigma^+\right)$ molecules. (b) Same as (a) but for $O_2 \left(X^3 \Sigma^+_g\right)$ (c) The variations of energy (in eV) in TW diatomic molecular potential, with respect to rotational ($\ell$) for $HF \left(X^1\Sigma^+\right)$ molecules. (d) Same as (c) but for $O_2 \left(X^3 \Sigma^+_g\right)$.}}
\label{fig8}
\end{figure}
\begin{figure}[!t]
\includegraphics[height=100mm,width=180mm]{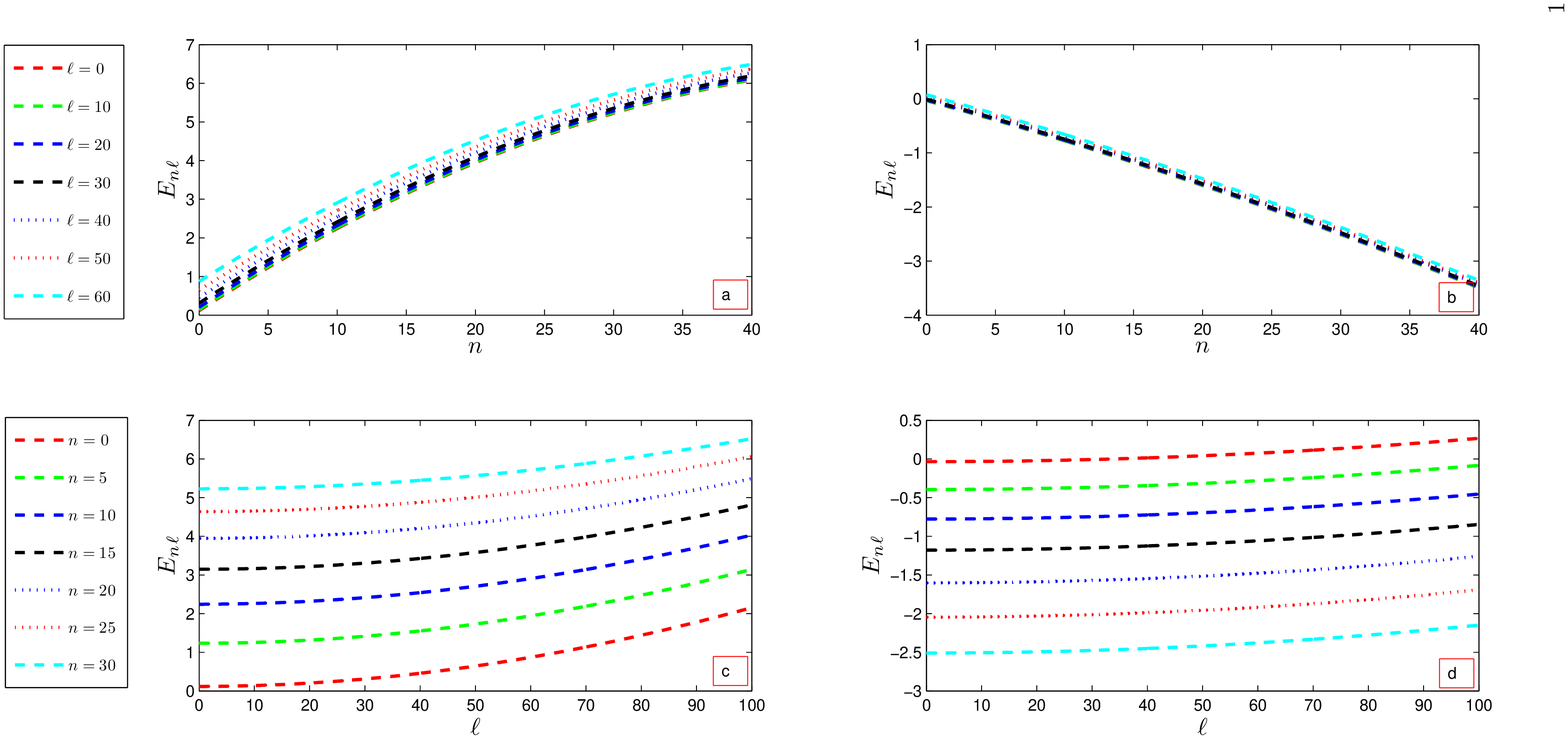}
\caption{{\protect\small (a) The variations of energy (in eV) in TW diatomic molecular potential, with respect to vibrational ($n$) for $NO \left(X^2 \Pi_r\right)$ molecules. (b) Same as (a) but for $Cl_2 \left(X^1\Sigma^+_g\right)$ (c) The variations of energy (in eV) in TW diatomic molecular potential, with respect to rotational ($\ell$) for $NO \left(X^2 \Pi_r\right)$ molecules. (d) Same as (c) but for $Cl_2 \left(X^1\Sigma^+_g\right)$.}}
\label{fig9}
\end{figure}
\begin{figure}[!t]
\includegraphics[height=100mm,width=180mm]{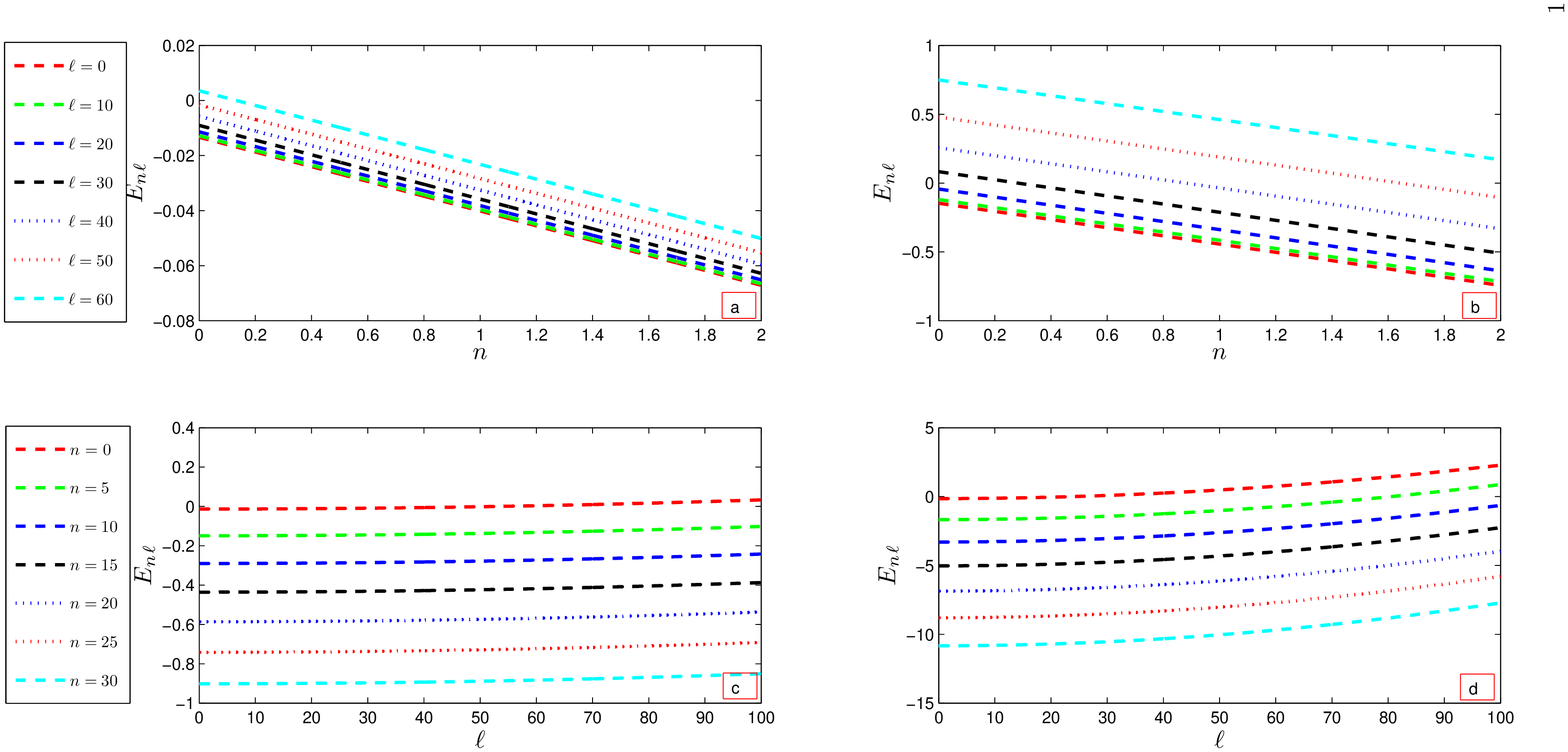}
\caption{{\protect\small (a) The variations of energy (in eV) in TW diatomic molecular potential, with respect to vibrational ($n$) for $I_2 \left(X\left(O_g^+\right)\right)$ molecules. (b) Same as (a) but for $N_2 \left(X^1\Sigma^+_g\right)$ (c) The variations of energy (in eV) in TW diatomic molecular potential, with respect to rotational ($\ell$) for $I_2 \left(X\left(O_g^+\right)\right)$ molecules. (d) Same as (c) but for $N_2 \left(X^1\Sigma^+_g\right)$.}}
\label{fig10}
\end{figure}
\begin{figure}[!t]
\includegraphics[height=100mm,width=180mm]{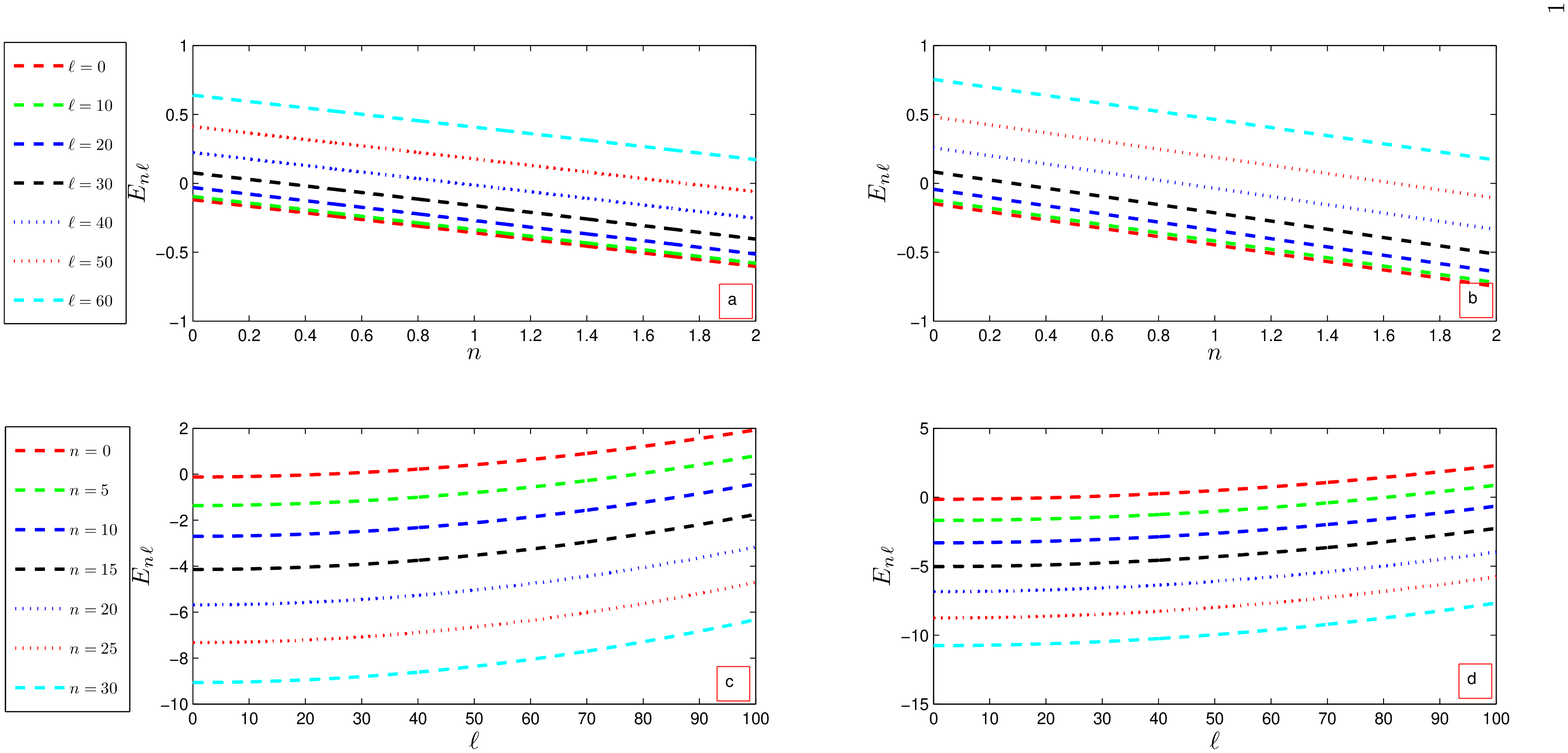}
\caption{{\protect\small (a) The variations of energy (in eV) in TW diatomic molecular potential, with respect to vibrational ($n$) for $O_2 ^+\left(X^2 \Pi_g\right)$ molecules. (b) Same as (a) but for $NO^+ \left(X^1 \Sigma^+\right)$ (c) The variations of energy (in eV) in TW diatomic molecular potential, with respect to rotational ($\ell$) for $O_2 ^+\left(X^2 \Pi_g\right)$ molecules. (d) Same as (c) but for $NO^+ \left(X^1 \Sigma^+\right)$.}}
\label{fig11}
\end{figure}
with $b_h=\beta(1-c_h)$, $r_e$ is the molecular bond length, $\beta$ is the Morse constant, $D$ is the potential well depth and $c_h$ is an optimization parameter obtained from ab initio or Rydberg-Klein-Rees (RKR).  intramolecular potentials. When the potential constant approaches zero, i.e. $c_h\rightarrow 0$, the TW potential reduces to the Morse potential \cite{BJ55}. The shape of this potential is shown in figure (\ref{fig1}) for different molecules. Inserting this potential into the Schr\"{o}dinger equation, the radial part of the Schr\"{o}dinger equation takes the following form:
\begin{equation}
\frac{d^2U_{n\ell}(r)}{dr^2}+\frac{2\mu}{\hbar^2}\left[E_{n\ell}-D\left[\frac{1-e^{-b_h(r-r_e)}}{1-c_he^{-b_h(r-r_e)}}\right]^2-\frac{\ell(\ell+1)\hbar^2}{2\mu r^2}\right]U_{n\ell}(r)=0,
\label{E5}
\end{equation}
in which $U_{n\ell}(0)=0$ and $\lim_{r\rightarrow\infty}U_{n\ell}(r)=0$. In order to solve the above equation for $\ell\neq0$ states, we need to apply the following approximation scheme \cite{T2} to deal with the centrifugal term:
\begin{equation}
\frac{1}{r^2}\approx\frac{\ell(\ell+1)}{r_e^2}\left(D_0+D_1\frac{e^{-\alpha x}}{1-c_he^{-\alpha x}}+D_2\frac{e^{-2\alpha x}}{\left(1-c_he^{-\alpha x}\right)^2}\right),
\label{E6}
\end{equation}
where $\alpha=b_hr_e$, $x=\frac{r-r_e}{r_e}$ and $D_i$ is the parameter of coefficients $i=0, 1, 2$ and they are obtained as follows:
\begin{subequations}
\begin{equation}
D_0=1-\frac{1}{\alpha}(1-c_h)(3+c_h)+\frac{3}{\alpha^2}(1-c_h)^2, \ \ \ \ \ \ \lim_{c_h\rightarrow 0} D_0=1-\frac{3}{\alpha}+\frac{3}{\alpha^2}
\end{equation}
\begin{equation}
D_1=\frac{2}{\alpha}(1-c_h)^2(2+c_h)-\frac{6}{\alpha^2}(1-c_h)^3, \ \ \ \ \ \ \lim_{c_h\rightarrow 0} D_1=\frac{4}{\alpha}-\frac{6}{\alpha^2}
\end{equation}
\begin{equation}
D_2=-\frac{1}{\alpha}(1-c_h)^3(1+c_h)+\frac{3}{\alpha^2}(1-c_h)^4, \ \ \ \ \ \ \lim_{c_h\rightarrow 0} D_2=-\frac{1}{\alpha}+\frac{3}{\alpha^2}.
\end{equation}
\label{E7}
\end{subequations}
By using approximation expression (\ref{E6}), the radial part of the Schr\"{o}dinger equation with the TW diatomic molecular potential reduces to
\begin{eqnarray}
&&\frac{d^2U_{n\ell}(z)}{dz^2}+\frac{1}{z}\frac{dU_{n\ell}(z)}{dz}+\frac{1}{z^2(1-c_hz)^2}\left\{\left[\frac{2\mu r_e^2}{\hbar^2\alpha^2}(E_{n\ell}-D)-\frac{\ell(\ell+1)}{\alpha^2}D_0\right]\right.\nonumber\\
&&\left.+\left[-2c_h\left(\frac{2\mu r_e^2E_{n\ell}}{\alpha^2\hbar^2}-\frac{\ell(\ell+1)}{\alpha^2}D_0\right)+\frac{4\mu r_e^2D}{\hbar^2\alpha^2}-\frac{\ell(\ell+1)}{\alpha^2}D_1\right]z\right.\nonumber\\
&&\left.+\left[c_h^2\left(\frac{2\mu r_e^2E_{n\ell}}{\alpha^2\hbar^2}-\frac{\ell(\ell+1)}{\alpha^2}D_0\right)+\frac{\ell(\ell+1)}{\alpha^2}\left(D_1c_h-D_2\right)-\frac{2\mu r_e^2D}{\hbar^2\alpha^2}\right]z^2\right\}U_{n\ell}(z)=0,
\label{E8}
\end{eqnarray}
where we have introduced a new transformation of the form $z=e^{-b_h(r-r_e) }$ $\in(e^\alpha, 0)$ which maintained the finiteness of the transformed wave functions on the boundary conditions. By using the following transformation:
\begin{equation}
U_{n\ell}(z)=z^p(1-c_hz)^qR_{n\ell}(z),
\label{E9}
\end{equation}
where
\begin{equation}
p=\sqrt{\left[\frac{2\mu r_e^2}{\hbar^2\alpha^2}(D-E_{n\ell})+\frac{\ell(\ell+1)}{\alpha^2}D_0\right]}\ \ \mbox{and} \ \ q=\frac{1}{2}\sqrt{1+\frac{4}{c_h^2}\left[\frac{\ell(\ell+1)}{\alpha^2}D_2+\frac{2\mu r_e^2D}{\hbar^2\alpha^2}(1-c_h)^2\right]},
\label{E10}
\end{equation}
equation (\ref{E8}) is transformed into a more convenient second-order homogeneous linear differential equation which the solution can be found by using asymptotic iteration method \cite{N1,N2}
\begin{equation}
R_{n\ell}''(z)+\left[\frac{(2p+1)-zc_h(2p+2q+1)}{z(1-c_hz)}\right]R_{n\ell}'(z)-\left[\frac{c_h(p+q)^2+\varsigma/c_h}{z(1-c_hz)}\right]R_{n\ell}(z)=0,
\label{E11}
\end{equation}
where $\varsigma = c_h^2\left(\frac{2\mu r_e^2E_{n\ell}}{\alpha^2\hbar^2}-\frac{\ell(\ell+1)}{\alpha^2}D_0\right)+\frac{\ell(\ell+1)}{\alpha^2}\left(D_1c_h-D_2\right)-\frac{2\mu r_e^2D}{\hbar^2\alpha^2}$ has been introduced for mathematical simplicity. The systematic procedure of the asymptotic iteration method begins now by re-writing equation (\ref{E11}) in the following form \cite{N1,N2}
\begin{equation}
R_{n\ell}''(z)-\lambda_0(z)R_{n\ell}'(z)-s_0(z)R_{n\ell}(z)=0,
\label{E12}
\end{equation}
where
\begin{equation}
\lambda_0(z)=\left[\frac{zc_h(2p+2q+1)-(2p+1)}{z(1-c_hz)}\right] \ \ \ \mbox{and}\ \ \ s_0(z)=\left[\frac{c_h(p+q)^2+\varsigma/c_h}{z(1-c_hz)}\right].
\label{E13}
\end{equation}
The primes of $R_{n\ell}(z)$ in equation (\ref{E12}) denotes derivatives with respect to z. The asymptotic aspect of the method for sufficiently  large $k$ is given as \cite{N1,N2}
\begin{equation}
\frac{s_k(z)}{\lambda_k(z)}=\frac{s_{k-1}(z)}{\lambda_{k-1}(z)}=\eta(z),
\label{E14}
\end{equation}
where 
\begin{equation}
\lambda_k(z)=\lambda_{k-1}'(z)+s_{k-1}(z)+\lambda_0(z)\lambda_{k-1}(z), \ \ \ \ s_k(z)=s_{k-1}'(z)+s_0(z)\lambda_{k-1}(z),
\label{E15}
\end{equation}
Equation (\ref{E15}) is refer as the recurrence relations \cite{N1,N21}. In accordance with asymptotic iteration method \cite{N1,N2}, the energy eigenvalues equation are obtained from the roots of the following equation:
\begin{equation}
\delta_k(z)= \left| 
\begin{array}{lr}
\lambda_k(z) & s_k(z) \\ 
\lambda_{k+1}(z) & s_{k+1}(z)%
\end{array}
\right|=0\ \ ,\ \ \ k=1, 2, 3.... 
\label{E16}
\end{equation}
Thus, we can easily obtain the following simple arithmetic progressions:
\begin{eqnarray}
&&\delta_0(z)= \left| 
\begin{array}{lr}
\lambda_0(z) & s_0(z) \\ 
\lambda_{1}(z) & s_{1}(z)%
\end{array}
\right|=0\ \ \ \Leftrightarrow\ \ \ \ p+q=\ \ \ 0 +\frac{1}{c_h}\sqrt{-\varsigma}\nonumber\\
&&\delta_1(z)= \left| 
\begin{array}{lr}
\lambda_1(z) & s_1(z) \\ 
\lambda_{2}(z) & s_{2}(z)
\end{array}
\right|=0\ \ \ \Leftrightarrow\ \ \ \ p+q=-1 +\frac{1}{c_h}\sqrt{-\varsigma}\nonumber\\
&&\delta_2(z)= \left| 
\begin{array}{lr}
\lambda_2(z) & s_2(z) \\ 
\lambda_{3}(z) & s_{3}(z)
\end{array}
\right|=0\ \ \ \Leftrightarrow\ \ \ \ p+q=-2 +\frac{1}{c_h}\sqrt{-\varsigma}\\
&&\delta_3(z)= \left| 
\begin{array}{lr}
\lambda_3(z) & s_3(z) \\ 
\lambda_{4}(z) & s_{4}(z)
\end{array}
\right|=0\ \ \ \Leftrightarrow\ \ \ \ p+q=-3 +\frac{1}{c_h}\sqrt{-\varsigma}\nonumber\\
\label{E17}
&&.... etc.\nonumber
\end{eqnarray}
By finding the nth term of the above progression and inserting the values of $p$, $q$ and $\varsigma$, the energy eigenvalues turns out as
\begin{eqnarray}
E_{n\ell}&=&\frac{\hbar^2\ell(\ell+1)D_0}{2\mu r_e^2}+\frac{\hbar^2}{2\mu r_e^2c_h^2}\left[\ell(\ell+1)(D_2-D_1c_h)+\frac{2\mu Dr_e^2}{\hbar^2}\right]\nonumber\\
&&-\frac{\alpha^2\hbar^2}{2\mu r_e^2}\left[\frac{(\delta+n)^2-\frac{\ell(\ell+1)}{\alpha^2c_h^2}(D_1c_h-D_2)+\frac{2\mu Dr_e^2}{\alpha^2\hbar^2}\left(\frac{1}{c_h^2}-1\right)}{2(\delta+n)}\right]^2\nonumber\\
&=&\frac{\hbar^2\ell(\ell+1)D_0}{2\mu r_e^2}+D-\frac{\alpha^2\hbar^2}{2\mu r_e^2}\left[\frac{(\delta+n)^2+\frac{\ell(\ell+1)}{\alpha^2c_h^2}(D_1c_h-D_2)+\frac{2\mu Dr_e^2}{\alpha^2\hbar^2}\left(1-\frac{1}{c_h^2}\right)}{2(\delta+n)}\right]^2\label{E18}\\
&&\mbox{with}\ \ \delta=\frac{1}{2}+\frac{1}{2}\sqrt{1+\frac{4}{c_h^2}\left(\frac{D_2\ell(\ell+1)}{\alpha^2}+\frac{2\mu Dr_e^2}{\alpha^2\hbar^2}(1-c_h)^2\right)}.\nonumber
\end{eqnarray}
\begin{table}[!t]
{\scriptsize
\caption{Model parameters of the diatomic molecules studied in the present work. } \vspace*{10pt}{
\begin{tabular}{cccccc}\hline\hline
{}&{}&{}&{}&{}&{}\\[-1.0ex]
Molecules(states)&$c_h$&$\mu/ 10^{-23}(g)$& $b_h(\AA^{-1})$&$r_e (\AA)$&$D(cm^{-1})$\\[2.5ex]\hline\hline
$H_2 \left(X^1\Sigma^+_g\right)$	&0.170066	&0.084	&1.61890	&0.741	&38318	\\[1ex]
$CO \left(X^1\Sigma^+\right)$	&0.149936	&1.146	&2.20481	&1.128	&90531	\\[1ex]
$HF \left(X^1\Sigma^+\right)$	&0.127772	&0.160	&1.94207	&0.917	&49382	\\[1ex]
$O_2 \left(X^3 \Sigma^+_g\right)$	& 0.027262	&1.337	&2.59103	&1.207	&42041	\\[1ex]
$NO \left(X^2 \Pi_r\right)$	&0.013727	&1.249	&2.71559	&1.151	&53341	\\[1ex]
$Cl_2 \left(X^1\Sigma^+_g\right)$	&-0.096988	&2.924	&2.20354	&1.987	&20276	\\[1ex]
$I_2 \left(X\left(O_g^+\right)\right)$	&-0.139013	&10.612	&2.12343	&2.666	&12547	\\[1ex]
$N_2 \left(X^1\Sigma^+_g\right)$	&-0.032325	&1.171	&2.78585	&1.097	&79885	\\[1ex]
$O_2 ^+\left(X^2 \Pi_g\right)$	&-0.019445	&1.337	&2.86987	&1.116	&54688	\\[1ex]
$NO^+ \left(X^1 \Sigma^+\right)$	&-0.029000	&1.239	&2.73349	&1.063	&88694	\\[1ex]
\hline\hline
\end{tabular}\label{Tab1}}
\vspace*{-1pt}}
\end{table}
\begin{table}[!t]
{\scriptsize
\caption{ \small Comparison of the bound-state energy eigenvalues $-(E_{n\ell}-D)(eV)$ of $H_2$ and $C0$ molecules for various $n$ and rotational $\ell$ quantum numbers in TW diatomic molecular potential. } \vspace*{10pt}{
\begin{tabular}{cccccccc}\hline\hline
{}&{}&{}&{}&{}&{}&{}&{}\\[-1.0ex]
$n$	&$\ell$	&Present  &N-U \cite{T2}	&Present	    &N-U \cite{T2}	\\[1ex]\hline\hline
0	&0	&4.481442828		&4.4815718267	&11.073782707	  &11.07370964	\\[1ex]
	&5	&4.266910238		&4.2658220403	&11.066682384	  &11.06659606	\\[1ex]
	&10	&3.737244885		&3.7336304360	&11.047753377	  &11.04763173	\\[1ex]
5	&0	&2.280571333	  &2.2669650930	&9.6323160070	  &9.629868985	\\[1ex]
	&5	&2.120961179		&2.1070072990	&9.6256074980	  &9.623148913	\\[1ex]
	&10	&1.725024443		&1.7105026080	&9.6077232750	  &9.084920084	\\[1ex]
7	&0	&1.628985073		&1.6130911000	&9.0881674810	  &9.084920084	\\[1ex]
	&5	&1.488262184		&1.4722799590	&9.0816125880	  &9.078354525	\\[1ex]
	&10	&1.138505495		&1.1225356530	&9.0641379900	  &9.060851440	\\[1ex]
\hline\hline
\end{tabular}\label{Tab2}}
\vspace*{-1pt}}
\end{table}
Let us now calculate the eigenfunction solution of this problem. Generally speaking, the differential equation we wish to solve should be transformed to the form that is suitable to apply AIM \cite{N1,N2}: 
\begin{equation}
y^{\prime \prime }(x)=2\left( \frac{\Lambda x^{N+1}}{1-bx^{N+2}}-\frac{m+1}{x}\right) y^{\prime }(x)-\frac{Wx^{N}}{1-bx^{N+2}}y(x),  \label{E19}
\end{equation}%
where $\Lambda$, $b$ and $m$ are constants. The general solution of equation (\ref{E19}) is found as {\cite{N1,N2}} 
\begin{equation}
y_{n}(x)=(-1)^{n}C_{2}(N+2)^{n}(\sigma )_{_{n}}{_{2}F_{1}(-n,t+n;\sigma ;bx^{N+2})},  
\label{E20}
\end{equation}
where 
\begin{equation}
(\sigma )_{_{n}}=\frac{\Gamma {(\sigma +n)}}{\Gamma {(\sigma )}}\ \ ,\ \
\sigma =\frac{2m+N+3}{N+2}\ \ and\ \ \ t=\frac{(2m+1)b+2\Lambda }{(N+2)b}.
\label{E21}
\end{equation}%
By comparing equation (\ref{E12}) with (\ref{E20}), we can deduce that $m=p-\frac{1}{2}$, $t=2(p+q)$ and $\sigma =2p+1$. It is therefore straightforward to show that radial wave functions can be written as
\begin{eqnarray}
U_{n\ell}(x)&=&(-1)^nC_2\frac{\Gamma(2p+1+n)}{\Gamma(2p+1)}e^{-p\alpha x}(1-c_he^{-\alpha x})^q\ _2F_1\left(-n, n+2(p+q); 2p+1, c_he^{-\alpha x}\right),\nonumber\\
&=&N_{n\ell}e^{-p\alpha x}(1-c_he^{-\alpha x})^qP_n^{\left(2p,\ \ 2q-1\right)}(1-2c_he^{-\alpha x}),
\label{E22}
\end{eqnarray}
where $N_{n\ell}$ is normalization constant to be calculated from the normalization condition
\begin{equation}
\int_0^\infty \left|U_{n\ell}(r)\right|^2dr=\int_0^{e^\alpha} \left|U_{n\ell}(z)\right|^2\frac{dz}{\alpha z}=1, \ \ \ z=e^{-\alpha x}.
\label{E23}
\end{equation}
By using equation (\ref{E23}) and the following two different forms of Jacobi polynomials \cite{BJ63,BJ64} 
\begin{eqnarray}
&&P_n^{(\alpha, \beta)}(s)=\frac{\Gamma(\alpha+n+1)}{n!\Gamma(\alpha+\beta+n+1)}\sum^n_{m=0}\left(\begin{matrix}n\\m\end{matrix}\right)\frac{\Gamma(\alpha+\beta+n+m+1)}{\Gamma(\alpha+m+1)}\left(\frac{s-1}{2}\right)^m,\nonumber\\
&&P_n^{(\alpha, \beta)}(s)=\frac{1}{2^n}\sum^n_{k=0}\left(\begin{matrix}n+\alpha\\k\end{matrix}\right)\left(\begin{matrix}n+\beta\\k\end{matrix}\right)\left(\begin{matrix}n+\beta\\n-k\end{matrix}\right)(s-1)^{n-k}(s+1)^k,
\label{E24}
\end{eqnarray}
then
\begin{eqnarray}
&&N_{n\ell}^2\left[\frac{(-c_h)^{m+n-k}W_{m,n,k}}{\alpha n!}\right]\int_0^{e^\alpha} z^{m+n-k+2p-1}(1-c_hz)^{2q+k}{dz}=1,\ \ \ \ \ \mbox{with}\\
&&W_{m,n,k}=\frac{\Gamma(2p+n+1)}{n!\Gamma(2p+2q+n)}\sum^n_{m=0}\sum^n_{k=0}\left(\begin{matrix}n\\m\end{matrix}\right)\frac{\Gamma(2p+2q+n+m)}{\Gamma(2p+m+1)}\left(\begin{matrix}n+2p\\k\end{matrix}\right)\left(\begin{matrix}n+2q-1\\k\end{matrix}\right)\left(\begin{matrix}n+2q-1\\n-k\end{matrix}\right).\nonumber
\label{E25}
\end{eqnarray}
Thus, the normalization constant can be found as
\begin{equation}
N_{n\ell}=\frac{e^{\alpha\left(\frac{k-m-n}{2}\right)-\alpha p}}{\sqrt{\left(\frac{(-c_h)^{m+n-k}W_{m,n,k}}{\alpha n!(m+n-k+2p)}\right)\ _2F_1\left(-2q-1, m+n-k+2p,  m+n-k+2p+1; c_h^2\right)}},
\label{E26}
\end{equation}
where we have utilized the the following integral
\begin{equation}
\int^w_0z^y(1-pz)^tdz=\frac{w^{1+y}}{1+y}\ _2F_1(-t, 1+y, 2+y; pw).
\label{E27}
\end{equation}

\subsection{Some expectation values for the TW diatomic molecular potential}
In this section we calculate some expectation values of the TW diatomic molecular potential within the framework of the Hellmann-Feynman theorem (HFT) \cite{BJ65, BJ66, BJ67, BJ68, BJ69, BJ70, BJ71,SD1,BJ72,BJ73}. In quantum mechanics, the Hellmann-Feynman theorem relates the derivative of the total energy with respect to a parameter, to the expectation value of the derivative of the Hamiltonian with respect to that same parameter. According to the theorem, once the spatial distribution of the electrons has been determined by solving the Schr\"{o}dinger equation, all the forces in the system can be calculated using classical electrostatics. This theorem find some of its application in the calculation of intramolecular forces in molecules.  Assuming that the Hamiltonian $H$ for a particular quantum system is a function of some parameters $q$, and denoting the eigenvalues and eigenfunctions of $H$𝐻, respectively, by $E_{n\ell}(q)$𝐸and $U_{n\ell}(q)$ theoretically we have that
\begin{equation}
\frac{\partial E_{n\ell}(q)}{\partial q}=\left\langle U_{n\ell}(q)\left|\frac{\partial H(q)}{\partial q}\right|U_{n\ell}(q)\right\rangle,
\label{E28}
\end{equation}
provided that the associated normalized eigenfunction $U_{n\ell}(q)$ is continuous with respect to the parameter $q$. The effective Hamiltonian of the TW diatomic molecular potential radial wave function is given by
\begin{equation}
H=-\frac{\hbar^2}{2\mu}\frac{d^2}{dr^2}+\frac{\hbar^2}{2\mu}\frac{\ell(\ell+1)}{r^2}+D\left[\frac{1-e^{-b_h(r-r_e)}}{1-c_he^{-b_h(r-r_e)}}\right]^2.
\label{E29}
\end{equation}
Having determined the effective Hamiltonian of the TW diatomic molecular potential, we can thus find the expectation values of the the following parameters
\begin{itemize}
	\item The expectation value of $V$\\ 
	by setting $q=D$, we can find
\begin{eqnarray}
&&\left\langle V\right\rangle_{n\ell}=\frac{D}{c_h^2}-\frac{\alpha^2\hbar^2D}{8\mu r_e^2}\left[\frac{2\eta\left[(\delta+n)^4-\zeta^2\right]+\frac{4\mu r_e^2}{\alpha^2\hbar^2}\left(\frac{1}{c_h^2}-1\right)\left[(\delta+n)^2+\zeta\right](\delta+n)}{(\delta+n)^3}\right]\nonumber\\
&&\mbox{with}\ \  \eta=\frac{2\mu r_e^2}{\alpha^2\hbar^2}\left(\frac{1}{c_h}-1\right)^2\left[1+\frac{4}{c_h^2}\left(\frac{D_2\ell(\ell+1)}{\alpha^2}+\frac{2\mu Dr_e^2}{\alpha^2\hbar^2}(1-c_h)^2\right)\right]^{-\frac{1}{2}}\\
&&\mbox{and}\ \ \ \zeta=\frac{\ell(\ell+1)}{\alpha^2c_h^2}(D_2-D_1c_h)-\frac{2\mu Dr_e^2}{\alpha^2\hbar^2}\left(1-\frac{1}{c_h^2}\right)\nonumber.
\label{E30}
\end{eqnarray}
\item The expectation value of momentum $\left\langle p^2\right\rangle_{n\ell}$. \\
	By setting $q=\mu$, we can find
	\begin{eqnarray}
	&&\left\langle p^2\right\rangle_{n\ell}=\frac{\hbar^2\ell(\ell+1)}{r_e^2}\left[D_0-\frac{D_1}{c_h}+\frac{D_2}{c_h^2}\right]-\frac{\alpha^2\hbar^2\xi}{r_e^2}\left[\xi-2\mu\rho\right],\nonumber\\
	&&\mbox{with}\ \ \ \xi=\frac{(\delta+n)^2-\frac{\ell(\ell+1)}{\alpha^2c_h^2}(D_1c_h-D_2)+\frac{2\mu Dr_e^2}{\alpha^2\hbar^2}\left(\frac{1}{c_h^2}-1\right)}{2(\delta+n)},\nonumber\\
	&&\rho=\frac{4\tau(n+\delta)[\delta+n-\xi]+\frac{4(\delta+n)Dr_e^2}{\alpha^2\hbar^2}\left[\frac{1}{c_h^2}-1\right]}{4(\delta+n)^2} \ \ \mbox{and}\\
&&\tau=\frac{2Dr_e^2}{\alpha^2\hbar^2c_h^2}(1-c_h)^2\left[1+\frac{4}{c_h^2}\left(\frac{D_2\ell(\ell+1)}{\alpha^2}+\frac{2\mu D r_e^2}{\alpha^2\hbar^2}(1-c_h)^2\right)\right]^{-\frac{1}{2}.\nonumber}
\label{E31}
\end{eqnarray}
It is pertinent to note that the expectation of the kinetic energy can be simply be obtain from the above result by using the relation $\left\langle T\right\rangle_{n\ell}=\frac{1}{2\mu}\left\langle p^2\right\rangle_{n\ell}$ as 
\begin{equation}
\left\langle T\right\rangle_{n\ell}=\frac{\hbar^2\ell(\ell+1)}{2\mu r_e^2}\left[D_0-\frac{D_1}{c_h}+\frac{D_2}{c_h^2}\right]-\frac{\alpha^2\hbar^2\xi}{2\mu r_e^2}\left[\xi-2\mu\rho\right].
\label{E32}
\end{equation}
\item The expectation value of $r^{-2}$. \\
By setting $q=\ell$, we can find
\begin{eqnarray}
&&\left\langle r^{-2}\right\rangle_{n\ell}=\frac{1}{r_e^2}\left[D_0-\frac{D1}{c_h}+\frac{D_2}{c_h}\right]-\frac{2\alpha^2\xi}{(2\ell+1)r_e^2}\left[\frac{4\varsigma(\delta+n)[\delta+n-\xi]-\frac{2\ell+1}{\alpha^2c_h^2}(D_1c_h-D_2)}{4(\delta+n)^2}\right]\nonumber\\
&&\mbox{with}\ \ \ \varsigma=\frac{D_2(2\ell+1)}{\alpha^2c_h^2}\left[1+\frac{c_h}{4}\left(\frac{D_2\ell(\ell+1)}{\alpha^2}+\frac{2\mu Dr_e^2}{\alpha^2\hbar^2}(1-c_h)^2\right)\right]^{-\frac{1}{2}}.
	\label{E33}
\end{eqnarray}
\end{itemize}
\begin{landscape}
\begin{table}[!t]
{\scriptsize
\caption{\small Expectation values of $\left\langle V\right\rangle_{n\ell} (eV)$, $\left\langle p^2\right\rangle_{n\ell} (eV)$, $\left\langle T\right\rangle_{n\ell}  (eV\AA^{-1})$, $\left\langle T/V\right\rangle_{n\ell} (eV\AA^{-1})$, $\left\langle r^{-2}\right\rangle_{n\ell} (eV\AA^{-1})$, $\left\langle V\right\rangle_{n\ell} (eV)$, $\left\langle p^2\right\rangle_{n\ell} (eV)$, $\left\langle T\right\rangle_{n\ell} (eV\AA^{-1})$, $\left\langle T/V\right\rangle_{n\ell}  (eV\AA^{-1})$, $\left\langle r^{-2}\right\rangle_{n\ell} (eV\AA^{-1})$	 corresponding to T-W potential with various $n$ and $\ell$ quantum numbers for  $H_2 \left(X^1\Sigma^+_g\right)$, $CO \left(X^1\Sigma^+\right)$, $HF \left(X^1\Sigma^+\right)$, $O_2 \left(X^3 \Sigma^+_g\right)$, $NO \left(X^2 \Pi_r\right)$ and $Cl_2 \left(X^1\Sigma^+_g\right)$ diatomic molecules.} \vspace*{10pt}{
\begin{tabular}{cccccccccccccc}\hline\hline
\multicolumn{2}{c}{}& \multicolumn{5}{c}{$H_2 \left(X^1\Sigma^+_g\right)$ Diatomic Molecule}&\multicolumn{1}{c}{}&\multicolumn{5}{c}{$CO \left(X^1\Sigma^+\right)$ Diatomic Molecule}\\[1.5ex]
{}&{}&{}&{}&{}&{}&{}&{}&{}&{}&{}&{}&{}&{}\\[-1.0ex]
$n$	&$\ell$	&$\left\langle V\right\rangle_{n\ell} $	&$\left\langle p^2\right\rangle_{n\ell} $	&$\left\langle T\right\rangle_{n\ell}  $	&$\left\langle T/V\right\rangle_{n\ell} $	&$\left\langle r^{-2}\right\rangle_{n\ell} $	&&&$\left\langle V\right\rangle_{n\ell} $	&$\left\langle p^2\right\rangle_{n\ell} $	&$\left\langle T\right\rangle_{n\ell} $	&$\left\langle T/V\right\rangle_{n\ell}  $	&$\left\langle r^{-2}\right\rangle_{n\ell}$	\\[2.5ex]\hline\hline
0	&0	&0.136646347	&11761824963.6	&112325124.629	&822013373.171	&17696441386&&	&0.075611988	&4076474146.97	&2853524.18576	&37739044.5780	&7815433017.7	\\[1ex]	
	&5	&0.144410929	&11652913581.6	&111285023.743	&770613585.229	&16930544843&&	&0.075632410	&4076661983.18	&2853655.67075	&37730592.8867	&7814265945.0	\\[1ex]	
	&10	&0.201431069	&11372661386.7	&108608622.520	&539185057.495	&15181051187&&	&0.075692046	&4077162943.39	&2854006.34195	&37705498.6986	&7811153289.6	\\[1ex]	
5	&0	&2.774220723	&11744033916.5	&112155220.588	&40427648.6215	&13156799314&&	&1.584869235	&4076193491.98	&2853327.72780	&18003553.0048	&7384185493.1	\\[1ex]	
	&5	&2.818977983	&11630287679.9	&111068946.966	&39400430.8071	&12620296343&&	&1.585263957	&4076378680.13	&2853457.35916	&17999887.9465	&7383030423.9	\\[1ex]	
	&10	&2.946710003	&11339677399.5	&108293626.294	&36750690.1540	&11396406805&&	&1.586321554	&4076872577.36	&2853803.08628	&17990066.8883	&7379949787.1	\\[1ex]	
7	&0	&4.494731690	&11729814715.1	&112019427.582	&24922383.6500	&11597034082&&	&1.881625750	&4075954419.26	&2853160.37735	&1516327.23848	&7215106777.5	\\[1ex]	
	&5	&3.785165467	&11614622325.2	&110919343.234	&29303697.3419	&11135810214&&	&2.185570465	&4076138565.08	&2853289.27908	&1305512.37069	&7213956498.7	\\[1ex]	
	&10	&3.928165719	&11320959571.8	&108114871.523	&27522991.4563	&10085139484&&	&2.186995973	&4076629682.30	&2853633.06020	&1304818.61669	&7210888638.4	\\[1ex]\hline
\multicolumn{2}{c}{}& \multicolumn{5}{c}{$HF \left(X^1\Sigma^+\right)$ Diatomic Molecule}&\multicolumn{1}{c}{}&\multicolumn{5}{c}{$O_2 \left(X^3 \Sigma^+_g\right)$ Diatomic Molecule}\\[1.5ex]\hline													
0	&0	&0.128298808	&10734402400.0	&53819464.4828	&419485304.047	&11659305959&&	&0.04899053	&12036391366.8	&7221815.28929	&14741247.5213	&6821023698	\\[1ex]	
	&5	&0.129283327	&10731490713.3	&53804866.0531	&416177919.471	&11588335479&&	&0.04900787	&12036676718.3	&7221986.49973	&14736381.1154	&6819660995	\\[1ex]	
	&10	&0.134115260	&10723642492.3	&53765517.1415	&400890376.990	&11401257947&&	&0.04905922	&12037437741.7	&7222443.11253	&14721887.3690	&6816026510	\\[1ex]	
5	&0	&2.657367932	&10729489031.8	&53794830.1496	&20243651.4349	&94490582249&&	&1.02823410	&12036342934.0	&7221786.22969	&7023484.46690	&6391758894	\\[1ex]	
	&5	&2.670751754	&10726160798.4	&53778143.2645	&20135957.2952	&93865980863&&	&1.02858175	&12036627749.4	&7221957.11846	&7021276.74194	&6390390585	\\[1ex]	
	&10	&2.707930317	&10717218863.8	&53733310.7612	&19842944.4155	&92219375660&&	&1.02951425	&12037387342.8	&7222412.87327	&7015359.79057	&6386741161	\\[1ex]	
7	&0	&3.321571334	&10725418673.9	&53774422.4479	&16189452.8344	&86193084922&&	&1.21330571	&12036301143.3	&7221761.15533	&5952136.46141	&6220935758	\\[1ex]	
	&5	&3.645331889	&10721939419.0	&53756978.3808	&14746799.4733	&85599535298&&	&1.41971453	&12036585744.4	&7221931.91553	&5086890.19019	&6219565251	\\[1ex]	
	&10	&3.692547139	&10712600971.5	&53710157.8663	&14545557.8072	&84034738438&&	&1.42099217	&12037344766.2	&7222387.32738	&5082636.96300	&6215909951	\\[1ex]\hline
\multicolumn{2}{c}{}& \multicolumn{5}{c}{$NO \left(X^2 \Pi_r\right)$ Diatomic Molecule}&\multicolumn{1}{c}{}&\multicolumn{5}{c}{$Cl_2 \left(X^1\Sigma^+_g\right)$ Diatomic Molecule}\\[1.5ex]\hline												 		
0	&0	&0.0590176	&26580624128.6	&17071992.8022	&289269519.638	&750378933&&	&-0.017349478	&1868181521.14	&512534.597566	&-29541787.8028	&2541386767.17	\\[1ex]
	&5	&0.0590366	&26581001395.9	&17072235.1104	&289180527.171	&750235856&&	&-0.017350980	&1868177966.36	&512533.622314	&-29539174.2895	&2541272947.27	\\[1ex]
	&10	&0.0590936	&26582007513.8	&17072881.3119	&288912527.108	&749854277&&	&-0.017354794	&1868168484.96	&512531.021097	&-29532532.6879	&2540969404.79	\\[1ex]
5	&0	&1.2390649	&26580599933.8	&17071977.2625	&13778113.8522	&705954452&&	&-0.364630615	&1868136832.41	&512522.337237	&-1405593.26659	&2626168878.47	\\[1ex]
	&5	&1.2394521	&26580976932.6	&17072219.3983	&13774004.9803	&705810846&&	&-0.364661440	&1868133492.16	&512521.420841	&-1405471.93814	&2626057663.46	\\[1ex]
	&10	&1.2404910	&26581982333.9	&17072865.1395	&13762989.9286	&705427857&&	&-0.364743463	&1868124582.90	&512518.976590	&-1405149.17629	&2625761068.01	\\[1ex]
7	&0	&1.4571502	&26580579009.2	&17071963.8232	&11715994.5647	&688228928&&	&-0.433577209	&1868098543.87	&512511.832796	&-1182054.36577	&2659649378.39	\\[1ex]
	&5	&1.7112425	&26580955900.5	&17072205.8900	&9976497.13001	&688085112&&	&-0.503935502	&1868095287.81	&512510.939498	&-1017016.93464	&2659539185.33	\\[1ex]
	&10	&1.7126703	&26581961015.6	&17072851.4474	&9968556.96476	&687701561&&	&-0.504047245	&1868086603.11	&512508.556855	&-101678674358	&2659245315.23	\\[1ex]\hline\hline
\end{tabular}\label{Tab5}}
\vspace*{-1pt}}
\end{table}
\end{landscape}
\begin{landscape}
\begin{table}[!t]
{\scriptsize
\caption{\small Expectation values of $\left\langle V\right\rangle_{n\ell} (eV)$, $\left\langle p^2\right\rangle_{n\ell} (eV)$, $\left\langle T\right\rangle_{n\ell}  (eV\AA^{-1})$, $\left\langle T/V\right\rangle_{n\ell} (eV\AA^{-1})$, $\left\langle r^{-2}\right\rangle_{n\ell} (eV\AA^{-1})$, $\left\langle V\right\rangle_{n\ell} (eV)$, $\left\langle p^2\right\rangle_{n\ell} (eV)$, $\left\langle T\right\rangle_{n\ell} (eV\AA^{-1})$, $\left\langle T/V\right\rangle_{n\ell}  (eV\AA^{-1})$, $\left\langle r^{-2}\right\rangle_{n\ell} (eV\AA^{-1})$	 corresponding to T-W potential with various $n$ and $\ell$ quantum numbers for  $H_2 \left(X^1\Sigma^+_g\right)$, $CO \left(X^1\Sigma^+\right)$, $HF \left(X^1\Sigma^+\right)$, $O_2 \left(X^3 \Sigma^+_g\right)$, $NO \left(X^2 \Pi_r\right)$ and $Cl_2 \left(X^1\Sigma^+_g\right)$ diatomic molecules.} \vspace*{10pt}{
\begin{tabular}{cccccccccccccc}\hline\hline
\multicolumn{2}{c}{}& \multicolumn{5}{c}{$I_2 \left(X\left(O_g^+\right)\right)$ Diatomic Molecule}&\multicolumn{1}{c}{}&\multicolumn{5}{c}{$N_2 \left(X^1\Sigma^+_g\right)$ Diatomic Molecule}\\[1.5ex]
{}&{}&{}&{}&{}&{}&{}&{}&{}&{}&{}&{}&{}&{}\\[-1.0ex]
$n$	&$\ell$	&$\left\langle V\right\rangle_{n\ell} $	&$\left\langle p^2\right\rangle_{n\ell} $	&$\left\langle T\right\rangle_{n\ell}  $	&$\left\langle T/V\right\rangle_{n\ell} $	&$\left\langle r^{-2}\right\rangle_{n\ell} $	&&&$\left\langle V\right\rangle_{n\ell} $	&$\left\langle p^2\right\rangle_{n\ell} $	&$\left\langle T\right\rangle_{n\ell} $	&$\left\langle T/V\right\rangle_{n\ell}  $	&$\left\langle r^{-2}\right\rangle_{n\ell}$	\\[2.5ex]\hline\hline												
0	&0	&-0.0066488341	&558513328.245	&42219.9566751	&-6349978.96475	&1409346961.10&&	&-0.07310727	&13906362690.4	&9526604.59497	&-130309948.586	&8348868372	\\[1ex]	
	&5	&-0.0066489510	&558513226.961	&42219.9490187	&-6349866.16967	&1409336874.82&&	&-0.07312401	&13906169383.6	&9526472.16944	&-130278306.256	&8347453937	\\[1ex]	
	&10	&-0.0066492578	&558512956.830	&42219.9285986	&-6349570.11271	&1409309977.40&&	&-0.07316254	&13905653924.6	&9526119.05237	&-130204870.585	&8343682339	\\[1ex]	
5	&0	&-0.1396815777	&558503274.986	&42219.1967145	&-302253.148981	&1433074557.92&&	&-1.53580225	&13906316966.1	&9526573.27137	&-6202994.73540	&8738662788	\\[1ex]	
	&5	&-0.1396839663	&558503189.934	&42219.1902851	&-302247.934415	&1433064691.37&&	&-1.53616158	&13906124198.4	&9526441.21515	&-6201457.80182	&8737250549	\\[1ex]	
	&10	&-0.1396903321	&558502963.077	&42219.1731362	&-302234.037972	&1433038379.92&&	&-1.53711365	&13905610176.6	&9526089.08264	&-6197387.60543	&8733484812	\\[1ex]	
7	&0	&-0.1662205103	&558494650.466	&42218.5447571	&-253991.187254	&1442464239.81&&	&-1.83508755	&13906277500.7	&9526546.23546	&-5191330.64548	&8893782600	\\[1ex]	
	&5	&-0.1929655280	&558494571.802	&42218.5388107	&-218787.983783	&1442454459.66&&	&-2.12204701	&13906084947.8	&9526414.32639	&-4489256.96815	&8892371246	\\[1ex]	
	&10	&-0.1929742157	&558494361.947	&42218.5229470	&-218778.051741	&1442428378.52&&	&-2.12335704	&13905571499.1	&9526062.58648	&-4486321.61574	&8888607862	\\[1ex]\hline
\multicolumn{2}{c}{}& \multicolumn{5}{c}{$O_2 ^+\left(X^2 \Pi_g\right)$ Diatomic Molecule}&\multicolumn{1}{c}{}&\multicolumn{5}{c}{$NO^+ \left(X^1 \Sigma^+\right)$ Diatomic Molecule}\\[1.5ex]\hline														
0	&0	&-0.0590531	&17376639579.0	&10425955.5513	&-176552214.046	&8073402239.2&&	&-0.0737188	&16182723972.9	&10477599.7280	&-142129276.765	&8888672445	\\[1ex]	
	&5	&-0.0590698	&17376605854.2	&10425935.3165	&-176501957.286	&8071902694.6&&	&-0.0737345	&16182434514.2	&10477412.3163	&-142096472.022	&8887171024	\\[1ex]	
	&10	&-0.0591083	&17376515914.2	&10425881.3526	&-176386080.341	&8067903742.1&&	&-0.0737702	&16181662688.7	&10476912.5933	&-142020932.481	&8883167721	\\[1ex]	
5	&0	&-1.2405016	&17376610367.6	&10425938.0245	&-8404614.73367	&8514294721.8&&	&-1.5485101	&16182688492.1	&10477576.7558	&-6766230.81490	&9276047969	\\[1ex]	
	&5	&-1.2408588	&17376576977.4	&10425917.9904	&-8402179.19267	&8512803533.9&&	&-1.5488481	&16182399470.1	&10477389.6268	&-6764633.42454	&9274543700	\\[1ex]	
	&10	&-1.2418058	&17376487929.6	&10425864.5619	&-8395728.67344	&8508826861.9&&	&-1.5497435	&16181628808.8	&10476890.6576	&-6760403.03289	&9270532816	\\[1ex]	
7	&0	&-1.4721570	&17376585124.8	&10425922.8788	&-7082072.68573	&8690035133.6&&	&-1.8589035	&16182657844.7	&10477556.9129	&-5636417.87371	&9430334769	\\[1ex]	
	&5	&-1.7140431	&17376551867.4	&10425902.9245	&-6082637.55124	&8688547269.4&&	&-2.1393993	&16182368997.2	&10477369.8969	&-4897341.93000	&9428829391	\\[1ex]	
	&10	&-1.7153485	&17376463175.1	&10425849.7092	&-6077977.57086	&8684579467.2&&	&-2.1406326	&16181598801.1	&10476871.2289	&-4894287.43115	&9424815522	\\[1ex]\hline\hline	
\end{tabular}\label{Tab6}}
\vspace*{-1pt}}
\end{table}
\end{landscape}
\section{Relativistic solutions of the Klein-Gordon equation and the Dirac equation for the TW diatomic molecular potential}\label{sec4}
The solutions of Klein-Gordon and Dirac equations are very significant in describing the nuclear shell structure. They are used to describe the particle dynamics in relativistic quantum mechanics. We obtain the approximate relativistic solutions of these equations in the presence of TW diatomic molecular potential via two eigensolution approaches.
\subsection{Relativistic treatment of the spin-zero particles subject to the TW diatomic molecular potential: Functional analysis approach}
The three dimensional Klein-Gordon equation with the scalar and vector potential can be written as follows $(\hbar=c=1)$\cite{BJ75}
\begin{equation}
\left\{\frac{1}{r^2}\frac{\partial}{\partial r}r^2\frac{\partial}{\partial r}+\frac{1}{r^2}\left[\frac{1}{\sin \theta}\frac{\partial}{\partial\theta}\left(\sin\theta\frac{\partial}{\partial\theta}\right)+\frac{1}{\sin^2\theta}\frac{\partial^2}{\partial\phi^2}\right]+(M+V_s)^2-(E_R-V_v)^2\right\}\psi(r,\theta,\phi)=0,
\label{E34}
\end{equation} 
where $M$ and $E_R$ represent the rest mass and energy of the spin-zero particle, respectively. The scalar potential $S(r)$ and vector one $V(r)$ are chosen as equal TW potentials as defined in equation (\ref{E4}). By taking $\psi(r,\theta,\phi)=\frac{R_{n\ell}(r)\Theta(\theta)e^{\pm im\phi}}{r}$ and substituting it into equation (\ref{E34}), we obtain the radial part of the Klein-Gordon equation as:
\begin{equation}
\left\{\frac{d^2}{dr^2}+\left(E_R-D\left[\frac{1-e^{-b_h(r-r_e)}}{1-c_he^{-b_h(r-r_e)}}\right]^2\right)^2-\left(M+D\left[\frac{1-e^{-b_h(r-r_e)}}{1-c_he^{-b_h(r-r_e)}}\right]^2\right)^2-\frac{\ell(\ell+1)}{r^2}\right\}R_{n\ell}(r)=0.
\label{E35}
\end{equation} 
 The above equation is solvable for $\ell=0$ \cite{T3} but it is not solvable for $\ell\neq0$. We therefore resort to use approximation expression (\ref{E6}) to deal with the centrifugal term. Then, we obtain the following equations
\begin{eqnarray}
&&\frac{d^2R_{n\ell}(z)}{dz^2}+\frac{1}{z}\frac{dR_{n\ell}(z)}{dz}+\frac{1}{z^2(1-c_hz)^2}\left\{\left[\frac{r_e^2}{\alpha^2}(\widetilde{E}_{n\ell}-\widetilde{D})-\frac{\ell(\ell+1)}{\alpha^2}D_0\right]\right.\nonumber\\
&&\left.+\left[-2c_h\left(\frac{r_e^2\widetilde{E}_{n\ell}}{\alpha^2}-\frac{\ell(\ell+1)}{\alpha^2}D_0\right)+\frac{2r_e^2\widetilde{D}}{\alpha^2}-\frac{\ell(\ell+1)}{\alpha^2}D_1\right]z\right.\nonumber\\
&&\left.+\left[c_h^2\left(\frac{r_e^2\widetilde{E}_{n\ell}}{\alpha^2}-\frac{\ell(\ell+1)}{\alpha^2}D_0\right)+\frac{\ell(\ell+1)}{\alpha^2}\left(D_1c_h-D_2\right)-\frac{r_e^2\widetilde{D}}{\alpha^2}\right]z^2\right\}R_{n\ell}(z)=0,
\label{E36}
\end{eqnarray}
where $\widetilde{E}_{n\ell}=E_R^2-M^2$ and $\widetilde{D}=(E_R+M)D$. It is worth to be noted that only the choice $V_s = +V_v$ produces a nontrivial nonrelativistic
limit with a potential function $2V_v(r)$, and not $V_v(r)$. Accordingly, it would be natural to scale the potential terms in equation (\ref{E35}) so that in the nonrelativistic limit the interaction potential becomes $V$, not $2V$ \cite{BJ76}. Thus, we have implement this modifications in equation (\ref{E36}) through which $\widetilde{D}$ becomes $(E_R+M)D$ and not $2(E_R+M)D$. Now, we express the solution to the above second order differential equation in terms of  the hypergeometric function as:
\begin{eqnarray}
R_{n\ell}(z)&=&\frac{\left[(\zeta+n)(\zeta+n-1)....\right]\left[(\beta+n)(\beta+n-1).....\right]\left[\Gamma(\gamma)\right]}{\left[\Gamma(\zeta)\right]\left[\Gamma(\beta)\right]\left[(\gamma+n)(\gamma+n-1)(\gamma+n-2).....\right]}\frac{(c_h)^n}{n!}z^u(1-c_hz)^v\label{E37}\\
&=&z^u(1-c_hz)^v+z^u(1-c_hz)^v\sum^\infty_{n=1}\frac{(\alpha)_n(\beta_n)(c_hz)^n}{n!(\gamma)_n}=z^u(1-c_hz)^v\ _2F_1(\zeta,\beta;\gamma;c_hz),\nonumber
\end{eqnarray}
where we have introduced the following notations to avoid mathematical complexity:
\begin{equation}
u=\sqrt{\left[\frac{r_e^2}{\alpha^2}(\widetilde{D}-\widetilde{E}_{n\ell})+\frac{\ell(\ell+1)}{\alpha^2}D_0\right]}\ \ \mbox{and} \ \ v=\frac{1}{2}\left[\sqrt{1+\frac{4}{c_h^2}\left[\frac{\ell(\ell+1)}{\alpha^2}D_2+\frac{r_e^2\widetilde{D}}{\alpha^2}(1-c_h)^2\right]}\right],
\label{E38}
\end{equation}
$\zeta$, $\beta$ and $\gamma$ are given by
\begin{eqnarray}
&\zeta=u+v-\sqrt{\left[\frac{r_e^2\widetilde{D}}{\alpha^2}-c_h^2\left(\frac{r_e^2\widetilde{E}_{n\ell}}{\alpha^2}-\frac{\ell(\ell+1)}{\alpha^2}D_0\right)-\frac{\ell(\ell+1)}{\alpha^2}\left(D_1c_h-D_2\right)\right]}\label{E39}\\
&\beta=u+v+\sqrt{\left[\frac{r_e^2\widetilde{D}}{\alpha^2}-c_h^2\left(\frac{r_e^2\widetilde{E}_{n\ell}}{\alpha^2}-\frac{\ell(\ell+1)}{\alpha^2}D_0\right)-\frac{\ell(\ell+1)}{\alpha^2}\left(D_1c_h-D_2\right)\right]}\ \ \mbox{and}\ \ \gamma=2u+1.\nonumber
\end{eqnarray}
Now by considering the finiteness of the solutions, it be readily seen from equation (\ref{E37}) that $R_{n\ell}(z)$ approaches infinity unless $\zeta$ is a negative integer. This implies that the wave function $R_{n\ell}(z)$ will not be finite everywhere unless we take
\begin{equation}
u+v-\sqrt{\left[\frac{r_e^2\widetilde{D}}{\alpha^2}-c_h^2\left(\frac{r_e^2\widetilde{E}_{n\ell}}{\alpha^2}-\frac{\ell(\ell+1)}{\alpha^2}D_0\right)-\frac{\ell(\ell+1)}{\alpha^2}\left(D_1c_h-D_2\right)\right]}=-n.
\label{E40}
\end{equation} 
Thus, the relativistic energy spectrum can be found by utilizing equation (\ref{E40}) as:
\begin{eqnarray}
&&E_R^2-M^2-\frac{\ell(\ell+1)D_0}{r_e^2}-\left[\frac{\ell(\ell+1)(D_2-D_1c_h)+{\tilde{D}r_e^2}}{r_e^2c_h^2}\right]\nonumber\\
&&+\frac{\alpha^2}{r_e^2}\left[\frac{(\delta_c+n)^2-\frac{\ell(\ell+1)}{\alpha^2c_h^2}(D_1c_h-D_2)+\frac{\tilde{D}r_e^2}{\alpha^2}\left(\frac{1}{c_h^2}-1\right)}{2(\delta_c+n)}\right]^2=0\label{E41}\\
&&\mbox{with}\ \ \delta_c=\frac{1}{2}+\frac{1}{2}\sqrt{1+\frac{4}{c_h^2}\left(\frac{D_2\ell(\ell+1)}{\alpha^2}+\frac{\tilde{D}r_e^2}{\alpha^2}(1-c_h)^2\right)}\nonumber
\end{eqnarray}
Hence, with equation (\ref{E40}), the expression for $\beta$ given by equation (\ref{E39}) can be rewritten as
\begin{equation}
\beta=2(v+u)+n,
\label{E42}
\end{equation} 
and the total radial wave functions can be written as
\begin{eqnarray}
R_{n\ell}(r)&=&e^{-ub_h(r-r_e)}\left(1-c_he^{-b_h(r-r_e)}\right)^v\ _2F_1\left(-n, n+2(v+u); 2u+1, c_he^{-b_h(r-r_e)}\right),\nonumber\\
&=&N_{n\ell}e^{-ub_h(r-r_e)}\left(1-c_he^{-b_h(r-r_e)}\right)^vP_n^{\left(2v,\ \ 2v-1\right)}(1-2c_he^{-b_h(r-r_e)})
\label{E43}
\end{eqnarray}
Before we proceed, let us first obtain the non-relativistic limit of equation (\ref{E41}) and eigenfunctions (\ref{E43}) of the TW diatomic molecular potential potential. de Souza Dutra et al \cite{BJ79} noted that there is possibility of obtaining approximate non-relativistic (NR) solutions from relativistic (R) ones. Very recently, Sun \cite{BJ79} proposed a meaningful approach for deriving the bound state solutions of NR Schr\"{o}dinger equation (SE) from the bound state of R equations. The essence of the approach was that, in NR limit, the SE may be derived from the R one when the energies of the potential  $V(r)$ are small compared to the rest mass $mc^2$, then the NR energy approximated as $E^{NR}\rightarrow E-mc^2$ and NR wave function is the $\psi^{NR} (r)\rightarrow\psi(r)$. That is, its NR energies, $E^{NR}$ can be determined by taking the NR limit values of the R eigenenergies $E$.  For this purpose we apply the following appropriate transformations.
\begin{equation}
M+E_R\rightarrow \frac{2\mu}{\hbar^2}\ \ \ \ \ \ \ \ \  \mbox{and}\ \ \ \ \ \ \ \ \ M-E_R\rightarrow-E_{n\ell},
\label{E44}
\end{equation}
then we can have the energy equation as
\begin{eqnarray}
E_{n\ell}=\frac{\hbar^2\ell(\ell+1)D_0}{2\mu r_e^2}+\frac{\hbar^2}{2\mu r_e^2c_h^2}\left[\ell(\ell+1)(D_2-D_1c_h)+\frac{2\mu Dr_e^2}{\hbar^2}\right]\nonumber\\
-\frac{\alpha^2\hbar^2}{2\mu r_e^2}\left[\frac{(\delta+n)^2-\frac{\ell(\ell+1)}{\alpha^2c_h^2}(D_1c_h-D_2)+\frac{2\mu Dr_e^2}{\alpha^2\hbar^2}\left(\frac{1}{c_h^2}-1\right)}{2(\delta+n)}\right]^2\label{E45}\\
\mbox{with}\ \ \delta=\frac{1}{2}+\frac{1}{2}\sqrt{1+\frac{4}{c_h^2}\left(\frac{D_2\ell(\ell+1)}{\alpha^2}+\frac{2\mu Dr_e^2}{\alpha^2\hbar^2}(1-c_h)^2\right)},\nonumber
\end{eqnarray}
and the eigenfunction as
\begin{eqnarray}
R_{n\ell}(r)=N_{n\ell}e^{-pb_h(r-r_e)}\left(1-c_he^{-b_h(r-r_e)}\right)^qP_n^{\left(2p,\ \ 2q-1\right)}(1-2c_he^{-b_h(r-r_e)}),
\label{E46}
\end{eqnarray}
where $N_{n\ell}$ is normalization constant. It is pertinent to note that the results we obtain in equations (\ref{E45}) and (\ref{E46}) as a non-relativistic limit solutions are exactly the same with the ones we obtain previously for Schr\"{o}dinger system. 

\subsection{Fermionic massive spin-$\frac{1}{2}$ particles interacting with TW diatomic molecular potential: Supersymmetry approach}
In spherical coordinates, the Dirac equation for fermionic massive spin$-\frac{1}{2}$ particles interacting with arbitrary scalar potential S(r), the time-component $V(r)$ of a four-vector potential and the tensor potential $U(r)$ can be expressed as \cite{BJ80, BJ81} 
\begin{equation}
\left[\vec{\alpha}.\vec{p}+\beta(M+S(r))-i\beta\vec{\alpha}.\hat{r}U(r)\right]\psi(\vec{r})=[E_r-V(r)]\psi(\vec{r}),
\label{E47}
\end{equation}
where $E_r$, $\vec{p}$ and $M$ denote the relativistic energy of the system, the momentum operator and mass of the particle respectively. $\alpha$ and $\beta$ are $4\times 4$ Dirac matrices. The eigenvalues of the spin-orbit coupling operator are $\kappa=\left(j+\frac{1}{2}\right)>0$ and $\kappa=-\left(j+\frac{1}{2}\right)<0$ for unaligned spin $j=\ell-\frac{1}{2}$ and the aligned spin $j=\ell+\frac{1}{2}$ respectively. The set $(H^2, K, J^2, J_Z)$ can be taken as the complete set of conservative quantities with $\vec{J}$ being the total angular momentum operator and $K=(\vec{\sigma}.\vec{L}+1)$ is the spin-orbit where $\vec{L}$ is the orbital angular momentum of the spherical nucleons that commutes with the Dirac Hamiltonian. Thus, the spinor wave functions can be classified according to their angular momentum $j$, the spin-orbit quantum number $\kappa$ and the radial quantum number $n$. Hence, they can be written as follows:
\begin{equation}
\psi_{n\kappa}(\vec{r})=
\left(\begin{array}{lr}     
f_{n\kappa}(\vec{r})\\      
g_{n\kappa}(\vec{r})
\end{array}\right)=\left(\begin{array}{lr}     
\frac{F_{n\kappa}({r})}{r}Y^\ell_{jm}(\theta,\phi)\\      
\frac{iG_{n\kappa}({r})}{r}Y^{\tilde{\ell}}_{jm}(\theta,\phi),
\end{array}\right),
\label{E48}
\end{equation}
where $F_{n\kappa}(\vec{r})$ and $G_{n\kappa}(\vec{r})$ are the radial wave functions of the upper- and lower-spinor components respectively and $Y^\ell_{jm}(\theta,\phi)$ and $Y^{\tilde{\ell}}_{jm}(\theta,\phi)$ are the spherical harmonic functions coupled to the total angular momentum $j$ and it's projection $m$ on the $z$ axis. The orbital angular
momentum quantum numbers $\ell$ and $\tilde{\ell}$ refer to the upper and lower components, respectively. The quasi-degenerate doublet structure can be expressed in terms of pseudospin
angular momentum $s = 1/2$ and pseudo-orbital angular momentum $\tilde{\ell}$, which is defined as $\tilde{\ell}=\ell+1$ for the aligned spin $j = \tilde{\ell}-1/2$ and $\tilde{\ell}=\ell-1$ for the unaligned spin $j = \tilde{\ell}+1/2$. Substitution of equation (\ref{E48}) into equation (\ref{E47}) yields the following coupled differential equations:
\begin{eqnarray}
\left(\frac{d}{dr}+\frac{\kappa}{r}\right)F_{n\kappa}(r)&=&(M+E_{n\kappa}-\Delta(r))G_{n\kappa}(r)\nonumber\\
\left(\frac{d}{dr}-\frac{\kappa}{r}\right)G_{n\kappa}(r)&=&(M-E_{n\kappa}+\sum(r))F_{n\kappa}(r)
\label{E49}
\end{eqnarray}
where $\Delta(r)=V(r)-S(r)$ and $\sum(r)=V(r)+S(r)$ are the difference and sum potentials respectively. On solving equation (\ref{E49}), we obtain the following Schr$\ddot{o}$diger-like differential equation with coupling to the $r^{-2}$ singular term and satisfying $F_{n\kappa}(r)$ and $F_{n\kappa}(r)$ respectively as:
\begin{equation}
\left[\frac{d^2}{dr^2}-\frac{\kappa(\kappa+1)}{r^2}+\frac{\frac{d\Delta(r)}{dr}\left(\frac{d}{dr}+\frac{\kappa}{r}\right)}{M+E_{n\kappa}-\Delta(r)}\right]F_{n\kappa}(r)=\left[\left(M+E_{n\kappa}-\Delta(r)\right)\left(M-E_{n\kappa}+\sum(r)\right)\right]F_{n\kappa}(r),
\label{E50}
\end{equation}
\begin{equation}
\left[\frac{d^2}{dr^2}-\frac{\kappa(\kappa-1)}{r^2}+\frac{\frac{d\sum(r)}{dr}\left(\frac{d}{dr}-\frac{\kappa}{r}\right)}{M-E_{n\kappa}+\sum(r)}\right]G_{n\kappa}(r)=\left[\left(M+E_{n\kappa}-\Delta(r)\right)\left(M-E_{n\kappa}+\sum(r)\right)\right]G_{n\kappa}(r).
\label{E51}
\end{equation}
The spin-orbit quantum number $\kappa$ is related to the orbital angular momentum quantum number $\ell$ via $\kappa(\kappa-1)=\tilde{\ell}(\tilde{\ell}+1)$ and $\kappa(\kappa+1) = \ell(\ell+1)$. 

\subsubsection{Spin symmetry solutions of the Dirac equation with TW diatomic molecular potential:}
\label{SSL}
In the spin symmetry limit, $\frac{d\Delta(r)}{dr}=0$ or $\Delta(r)=C_s=$ constant. We take
\begin{equation}
\sum(r)=D\left[\frac{1-e^{-b_h(r-r_e)}}{1-c_he^{-b_h(r-r_e)}}\right]^2,
\label{E52}
\end{equation}
and as a consequence, we can rewrite equation (\ref{E50}) as
\begin{equation}
\frac{d^2F_{n\kappa}(x)}{dx^2}=\left[V_{eff}(x)-\widetilde{E}^s\right]F_{n\kappa}(x),
\label{E53}
\end{equation}
where we have approximate the centrifugal term by approximation (\ref{E6}) and have also introduced the following parameters for convenience:
\begin{eqnarray}
&&V_{eff}(x)=\frac{D_1\kappa(\kappa+1)+2{E}_ar_e^2(c_h-1)}{\left(e^{\alpha x}-c_h\right)}+\frac{D_2\kappa(\kappa+1)+2{E}_ar_e^2(c_h-1)^2}{\left(e^{\alpha x}-c_h\right)^2}\ \ \ \mbox{and}\nonumber\\
&&\widetilde{E}^s=-D_0\kappa(\kappa+1)-(M+E_r-C_s)(M-E_r+D)r_e^2\ \ \ \ \ \ {E}_a=D(M+E_r-C_s)
\label{E54}
\end{eqnarray}
Now, let us apply the basic concepts of the supersymmetric quantum mechanics formalism and shape invariance technique to obtain approximate relativistic solution to the above equation (\ref{E54}). For a good SUSY, the ground-state function for the upper component $F_{nk}(r)$ can be written in the form of 
\begin{equation}
F_{0k}(r)=exp\left(\int{W(r)}dr\right),
\label{E55}
\end{equation} 
where $W(r)$ is named as a superpotential in supersymmetric quantum mechanics \cite{BJ43,BJ44,BJ45}. Now by Substituting equation (\ref{E55}) into equation (\ref{E53}),
we obtain the following equation for $W(r)$:
\begin{equation}
W^2(r)-W'(r)=\frac{D_1\kappa(\kappa+1)+2{E}_ar_e^2(c_h-1)}{\left(e^{\alpha x}-c_h\right)}+\frac{D_2\kappa(\kappa+1)+2{E}_ar_e^2(c_h-1)^2}{\left(e^{\alpha x}-c_h\right)^2}-\widetilde{E}^s_0
\label{E56}
\end{equation}
where the $\widetilde{E}^s_0$ denotes the ground-state energy. Since it is required that the superpotential should be made compatible with the right hand side of non-linear Riccati equation (\ref{E56}), therefore, we take the superpotential $W(r)$ as
\begin{equation}
W(r)=\mathcal{M}_a+\frac{\mathcal{M}_b}{e^{ax}-c_h}
\label{E57}
\end{equation}
where $\mathcal{M}_a$ and $\mathcal{M}_b$ are two parametric constant to be determine later. We now construct a pair of supersymmetric partner potentials $V_{-}(r)$ and $V_{+}(r)$ as follows
\begin{equation}
V_{-}(r)=W^2(r)-W'(r)=\mathcal{M}_a^2+\frac{2\mathcal{M}_a\mathcal{M}_b+\alpha\mathcal{M}_b}{e^{\alpha x}-c_h}e^{\alpha x}-\frac{\left(2\mathcal{M}_b\mathcal{M}_a-\mathcal{M}_b^2\right)}{\left(e^{\alpha x}-c_h\right)^2},
\label{E58}
\end{equation}
\begin{equation}
V_{+}(r)=W^2(r)+W'(r)=\mathcal{M}_a^2+\frac{2\mathcal{M}_a\mathcal{M}_b-\alpha\mathcal{M}_b}{e^{\alpha x}-c_h}e^{\alpha x}-\frac{\left(2\mathcal{M}_b\mathcal{M}_a-\mathcal{M}_b^2\right)}{\left(e^{\alpha x}-c_h\right)^2}.
\label{E59}
\end{equation}
On comparing equation (\ref{E58}) with (\ref{E56}), we can establish the following relationship between the parametric constants an other variables
\begin{subequations}
\begin{eqnarray}
e^{2ax}\ &:&\ \mathcal{M}_a^2=-\widetilde{E}^s_0,
\label{E60a}\\
e^{ax}\ &:&\ \alpha\mathcal{M}_b-2\mathcal{M}_a^2c_h+2\mathcal{M}_a\mathcal{M}_b=M+2\widetilde{E}^s_0c_h,
\label{E60b}\\
constant\ &:&\ \mathcal{M}_a^2c_h^2-2\mathcal{M}_a\mathcal{M}_bc_h+\mathcal{M}_b^2=\kappa(\kappa+1)(D_2-D_1c_h)+{E}_ar_e^2(1-c_h^2).
\label{E60c}
\end{eqnarray}
\end{subequations}
Since our interest lies only in the bound-state solutions; which demands that the radial part of the wave function 	$F_{n\kappa}$ must satisfy the boundary conditions
\begin{equation}
\frac{F_{n\kappa}(r)}{r}=\left\{\begin{matrix}0,\ \ \ \ r\rightarrow\infty\\ \infty, \ \ \ \ r\rightarrow0\end{matrix}\right..
\label{E61}
\end{equation}
 By considering these regularity conditions and their consequences (i.e. the restriction conditions $\mathcal{M}_a>0,\mathcal{M}_b>0$ ), we can solve equations (\ref{E60b}) and (\ref{E60c}) to have
\begin{subequations}
\begin{eqnarray}
\mathcal{M}_b&=&\alpha c_h\left[-\frac{1}{2}\pm\sqrt{1+\frac{4}{\alpha^2c_h^2}\left(D_2\kappa(\kappa+1)+E_ar_e^2(c_h-1)^2\right)}\right],\\
\mathcal{M}_a&=&\frac{\kappa(\kappa+1)\left(D_1-\frac{D_2}{c_h}\right)+{E}_ar_e^2\left(c_h-\frac{1}{c_h}\right)+\frac{\mathcal{M}_b^2}{c_h}}{2\mathcal{M}_b}.
\label{E62}
\end{eqnarray}
\end{subequations}
From these relations , we can obtain the supersymmetric partner potentials $V_{-}(r)$ and $V_{+}(r)$ as
\begin{equation}
V_{-}(r)=\left[\frac{\kappa(\kappa+1)\left(D_1-\frac{D_2}{c_h}\right)+{E}_ar_e^2\left(c_h-\frac{1}{c_h}\right)+\frac{\mathcal{M}_b^2}{c_h}}{2\mathcal{M}_b}\right]^2+\frac{2\mathcal{M}_a\mathcal{M}_b+\alpha\mathcal{M}_b}{e^{\alpha x}-c_h}e^{\alpha x}-\frac{\left(2\mathcal{M}_b\mathcal{M}_a-\mathcal{M}_b^2\right)}{\left(e^{\alpha x}-c_h\right)^2},
\label{E63}
\end{equation}
\begin{equation}
V_{+}(r)=\left[\frac{\kappa(\kappa+1)\left(D_1-\frac{D_2}{c_h}\right)+{E}_ar_e^2\left(c_h-\frac{1}{c_h}\right)+\frac{\mathcal{M}_b^2}{c_h}}{2\mathcal{M}_b}\right]^2+\frac{2\mathcal{M}_a\mathcal{M}_b-\alpha\mathcal{M}_b}{e^{\alpha x}-c_h}e^{\alpha x}-\frac{\left(2\mathcal{M}_b\mathcal{M}_a-\mathcal{M}_b^2\right)}{\left(e^{\alpha x}-c_h\right)^2}.
\label{E64}
\end{equation}
With the help of equations (\ref{E63}) and (\ref{E64}), we get the following relationship, which is satisfied by the partner potentials $V_{-}(r)$ and $V_{+}(r)$, 
\begin{equation}
V_{+}(r,a_0) = V_{-}(r,a_1)+ R(a_1),
\label{E65}
\end{equation}
where $a_0 = \mathcal{M}_b$, $a_1$ is a function of $a_0$, i.e., $a_1=h(a_0) =a_0 −c_h\alpha$, and the reminder $R(a_1)$ is independent of r, $R(a_1)=\left[\frac{\chi_m+\frac{a_0^2}{c_h}}{2a_0}\right]^2-\left[\frac{\chi_m+\frac{a_1^2}{c_h}}{2a_1}\right]^2$. Consequently, $a_n = f(a_0) = a_0 -nc_h\alpha$ with the remainder $R(a_n)=\left[\frac{\chi_m+\frac{a_{n-1}^2}{c_h}}{2a_{n-1}}\right]^2-\left[\frac{\chi_m+\frac{a_n^2}{c_h}}{2a_n}\right]^2$. We see that the shape invariance holds via a mapping of the form $\mathcal{M}_b\rightarrow\mathcal{M}_b-\alpha$. Thus, the energy spectra of the potential $V_{-}(r)$ can be determined by using the shape invariance approach and the following results can be obtained
\begin{subequations}
\begin{eqnarray}
\widetilde{E}_0^{s (-)}=0,
\label{E66a}
\end{eqnarray}
\begin{eqnarray}
\widetilde{E}_n^{s (-)}&=&\sum_{k=1}^nR(a_k)=R(a_1)+R(a_2)+R(a_3)+.......+R(a_n)\nonumber\\
&=&\left[\frac{\chi_mc_h+{a_0^2}}{2a_0c_h}\right]^2-\left[\frac{\chi_mc_h+{a_1^2}}{2a_1c_h}\right]^2+...+\left[\frac{\chi_mc_h+{a_{n-1}^2}}{2a_{n-1}c_h}\right]^2-\left[\frac{\chi_mc_h+{a_n^2}}{2a_nc_h}\right]^2\\
&=&\left[\frac{\chi_mc_h+{a_0^2}}{2a_0c_h}\right]^2-\left[\frac{\chi_mc_h+{a_n^2}}{2a_nc_h}\right]^2=\left[\frac{\chi_mc_h+{\mathcal{M}_b^2}}{2\mathcal{M}_bc_h}\right]^2-\left[\frac{\chi_mc_h+{\left(\mathcal{M}_b-nc_h\alpha\right)^2}}{2\left(\mathcal{M}_b-nc_h\alpha\right)c_h}\right]^2\nonumber\\
&&\mbox{with}\ \ \ \ \chi_m=\kappa(\kappa+1)\left(D_1-\frac{D_2}{c_h}\right)+{E}_ar_e^2\left(c_h-\frac{1}{c_h}\right)\nonumber
\label{E66b}
\end{eqnarray}
\end{subequations}
where $n=0, 1, 2, 3,....$ denotes the quantum numbers. We can obtain the energy levels of system as
\begin{equation}
\widetilde{E}^s=\widetilde{E}_n^{s (-)}+\widetilde{E}_0^s=-\left[\frac{\chi_mc_h+{\left(\mathcal{M}_b-nc_h\alpha\right)^2}}{2\left(\mathcal{M}_b-nc_h\alpha\right)c_h}\right]^2
\label{E67}
\end{equation}
Thus, the relativistic energy spectrum can be obtain directly from equation (\ref{E67}) as follows:
\begin{equation}
(M+E-C_s)(M-E+D)r_e^2+\kappa(\kappa+1)D_0-\alpha^2\left[\frac{\frac{\kappa(\kappa+1)}{\alpha^2c_h^2}(D_1c_h-D_2)+\frac{D(M+E-C_{s})r_e^2}{\alpha^2}\left(1-\frac{1}{c_h^2}\right)+\delta_s^2}{2\delta_s}\right]^2=0
\label{E68}
\end{equation}
with
\begin{equation}
\delta_s=n+\frac{1}{2}+\frac{1}{2}\sqrt{1+\frac{4}{\alpha^2c_h^2}\left[D_2\kappa(\kappa+1)+D(M+E-C_{s})r_e^2(c_h-1)^2\right]}
\label{E69}
\end{equation}
\subsubsection{Pseudospin symmetry solutions of the Dirac equation with the TW diatomic molecular potential}
\label{PSL}
In the pseudospin symmetry limit, $\frac{d\sum(r)}{dr}=0$ or $\sum(r)=C_{ps}=$ constant. By taking
\begin{equation}
\Delta(r)=D\left[\frac{1-e^{-b_h(r-r_e)}}{1-c_he^{-b_h(r-r_e)}}\right]^2,
\label{E70}
\end{equation}
and then substitute it into equation (\ref{E51}), we can find:
\begin{equation}
\frac{d^2G_{n\kappa}(x)}{dx^2}=\left[\bar{V}_{eff}(x)-\widetilde{E}^{ps}\right]G_{n\kappa}(x),
\label{E71}
\end{equation}
where we have introduced the following parameters for mathematical simplicity
\begin{eqnarray}
&&\bar{V}_{eff}(x)=\frac{D_1\kappa(\kappa-1)+2{E}_br_e^2(c_h-1)}{\left(e^{\alpha x}-c_h\right)}+\frac{D_2\kappa(\kappa+1)+2{E}_br_e^2(c_h-1)^2}{\left(e^{\alpha x}-c_h\right)^2}\ \ \ \mbox{and}\nonumber\\
&&\widetilde{E}^{ps}=-D_0\kappa(\kappa-1)-(M-E_r+C_{ps})(M-E_r+D)r_e^2\ \ \ \ \ \ {E}_b=D(E_r-M+C_s).
\label{E72}
\end{eqnarray}
By employing the same procedure of solving equation (\ref{E50}), we obtain the relativistic energy spectrum equation for the TW potential with pseudospin symmetry within the framework of Dirac theory as
\begin{equation}
(M-E_r+C_{ps})(M+E_r-D)r_e^2+\kappa(\kappa-1)D_0-\alpha^2\left[\frac{\frac{\kappa(\kappa-1)}{\alpha^2c_h^2}(D_1c_h-D_2)+\frac{D(E_r-M-C_{ps})r_e^2}{\alpha^2}\left(1-\frac{1}{c_h^2}\right)+\delta_{ps}^2}{2\delta_{ps}}\right]^2=0
\label{E73}
\end{equation}
with
\begin{equation}
\delta_{ps}=n+\frac{1}{2}+\frac{1}{2}\sqrt{1+\frac{4}{\alpha^2c_h^2}\left[D_2\kappa(\kappa-1)+D(E-M+C_{ps})r_e^2(c_h-1)^2\right]}.
\label{E74}
\end{equation}
Let us remark that a careful inspection to our present spin-symmetric solution shows that it can be easily recovered by knowing the relationship between the present set of parameters ($\bar{V}_{eff}$, $\widetilde{E}^{ps}$) and the previous set of parameters (${V}_{eff}$, $\widetilde{E}^{s}$). This tells us that the positive energy solution for spin symmetry (negative energy solution for pseudospin symmetry) can be obtained directly from those of the negative energy solution for pseudospin symmetry (positive energy
solution for spin symmetry) by performing the following parametric mappings:
\begin{equation}
\kappa\rightarrow\kappa-1, V(r)\rightarrow-V(r), E_r\rightarrow-E_r
\label{E75}
\end{equation}
\begin{landscape}
\begin{table}[!t]
{\scriptsize
\caption{The positive-energy degenerate states in units of $fm^{-1}$ of the spin-symmetry TW diatomic molecular potential for various values of $n$ and $b_h$. We use the following parameters: 
$ch=0.027262$, $M=5fm^{-1}$, $r_e=0.8$, $D=15fm^{-1}$.} \vspace*{10pt}{
\begin{tabular}{cccccccccccccc}\hline\hline
{}&{}&{}&{}&{}&{}&{}&{}&{}&{}&{}&{}&{}&{}\\[-1.0ex]
$\ell$&$n$,$\kappa<0$,$\kappa>0$&$b_h$&Degenerate states&$E_{n\kappa}(C_s=5fm^{-1})$&$E_{n\kappa}(C_s=0)$&&&$\ell$&$n$,$\kappa<0$,$\kappa>0$&$b_h$&Degenerate states&$E_{n\kappa}(C_s=5fm^{-1})$&$E_{n\kappa}(C_s=0)$	\\[2.5ex]\hline\hline
1	&0,\ \ \ \ -2,\ \ \ \ 1	&0.005	&$(0p_{1/2},\ \ 0p_{3/2})$	& 5.98591124	&-5.55887152	&&&1	&1,\ \ \ \ -2,\ \ \ \ 1	&0.005	&$(1p_{1/2},\ \ 1p_{3/2})$	&-1.94850126	& 6.20818681	\\[1ex]
	&	&	&	&-0.98583105	& 5.55895166	&&&	&	&	&	& 6.94868519	&-6.20800286	\\[1ex]
	&	&0.010	&	&-0.98442117	& 5.55831792	&&&	&	&0.010	&	& 6.94619996	&-6.20585892	\\[1ex]
	&	&	&	& 5.98474242	&-5.55799675	&&&	&	&	&	&-1.94546281	& 6.20659593	\\[1ex]
	&	&0.015	&	& 5.98370835	&-5.55711137	&&&	&	&0.015	&	& 6.94398591	&-6.20366597	\\[1ex]
	&	&	&	&-0.98298407	& 5.55783519	&&&	&	&	&	&-1.94232405	& 6.20532708	\\[1ex]
	&	&0.020	&	&-0.98151969	& 5.55750391	&&&	&	&0.020	&	&-1.93908461	&-6.20142391	\\[1ex]
	&	&	&	& 5.98280972	&-5.55621533	&&&	&	&	&	& 6.94204459	& 6.20438152	\\[1ex]\hline
2	&0,\ \ \ \ -3,\ \ \ \ 2	&0.005	&$(0d_{3/2},\ \ 0d_{5/2})$	&-1.87156670	& 6.15315790	&&&2	&1,\ \ \ \ -3,\ \ \ \ 2	&0.005	&$(1d_{3/2},\ \ 1d_{5/2})$	&-3.18687140	& 7.14788910	\\[1ex]
	&	&	&	& 6.87162480	&-6.15309970	&&&	&	&	&	& 8.18698960	&-7.14777100	\\[1ex]
	&	&0.010	&	& 6.86965028	& 6.15181620	&&&	&	&0.010	&	& 8.18314007	&-7.14445173	\\[1ex]
	&	&	&	&-1.86941685	&-6.15158281	&&&	&	&	&	&-3.18266653	& 7.14492526	\\[1ex]
	&	&0.015	&	& 6.86776253	& 6.15057742	&&&	&	&0.015	&	& 8.17944370	& 7.14215024	\\[1ex]
	&	&	&	&-1.86723613	&-6.15005114	&&&	&	&	&	&-3.17837588	&-7.14108260	\\[1ex]
	&	&0.020	&	& 6.86596228	&-6.14850468	&&&	&	&0.020	&	& 8.17590152	&-7.13766336	\\[1ex]
	&	&	&	&-1.86502432	& 6.14944223	&&&	&	&	&	&-3.17399907	& 7.13956523	\\[1ex]\hline
3	&0,\ \ \ \ -4,\ \ \ \ 3	&0.005	&$(0f_{5/2},\ \ 0f_{7/2})$	&-2.80058770	&-6.84443390	&&&3	&1,\ \ \ \ -4,\ \ \ \ 3	&0.005	&$(1f_{5/2},\ \ 1f_{5/2})$	&-4.34151230	&-8.09669090	\\[1ex]
	&	&	&	& 7.80063710	& 6.84448350	&&&	&	&	&	& 9.34160420	& 8.09678290	\\[1ex]
	&	&0.010	&	& 7.79803949	& 6.84251594	&&&	&	&0.010	&	& 9.33678307	&-8.09240555	\\[1ex]
	&	&	&	&-2.79784121	&-6.84231764	&&&	&	&	&	&-4.33641500	& 8.09277362	\\[1ex]
	&	&0.015	&	&-2.79506125	&-6.84018280	&&&	&	&0.015	&	&-4.33123946	&-8.08806926	\\[1ex]
	&	&	&	& 7.79550842	& 6.84062992	&&&	&	&	&	& 9.33206949	& 8.08889924	\\[1ex]
	&	&0.020	&	&-2.79224758	&-6.83802932	&&&	&	&0.020	&	& 9.32746431	&-8.08368196	\\[1ex]
	&	&	&	& 7.79304449	& 6.83882605	&&&	&	&	&	&-4.32598540	& 8.08516068	\\[1ex]\hline
4	&0,\ \ \ \ -5,\ \ \ \ 4	&0.005	&$(0g_{7/2},\ \ 0g_{9/2})$	& 8.75671840	&-7.60894320	&&&4	&1,\ \ \ \ -5,\ \ \ \ 4	&0.005	&$(1g_{7/2},\ \ 1g_{9/2})$	& 10.4562481	& 9.05826490	\\[1ex]
	&	&	&	&-3.75667370	& 7.60898800	&&&	&	&	&	&-5.45617040	&-9.05818720	\\[1ex]
	&	&0.010	&	& 8.75358668	&-7.60626790	&&&	&	&0.010	&	& 10.4506344	& 9.05338094	\\[1ex]
	&	&	&	&-3.75340755	& 7.60644708	&&&	&	&	&	&-5.45032368	&-9.05307026	\\[1ex]
	&	&0.015	&	& 8.75050995	& 7.60397470	&&&	&	&0.015	&	& 10.4451040	& 9.04860217	\\[1ex]
	&	&	&	&-3.75010591	&-7.60357068	&&&	&	&	&	&-5.44440334	&-9.04790158	\\[1ex]
	&	&0.020	&	&-3.74676852	&-7.60085145	&&&	&	&0.020	&	& 10.4396574	&-9.04268101	\\[1ex]
	&	&	&	& 8.74748866	& 7.60157143	&&&	&	&	&	&-5.43840894	& 9.04392931	\\[1ex]\hline\hline

\end{tabular}\label{Tab7}}
\vspace*{-1pt}}
\end{table}
\end{landscape}
\begin{landscape}
\begin{table}[!t]
{\scriptsize
\caption{The energy degenerate states in units of $fm^{-1}$ of the pseudospin-symmetry TW diatomic molecular potential for various values of $n$ and $b_h$. We use the following parameters: 
$ch=0.027262$, $M=5fm^{-1}$, $r_e=0.8$, $D=15fm^{-1}$.} \vspace*{10pt}{
\begin{tabular}{cccccccccccccc}\hline\hline
{}&{}&{}&{}&{}&{}&{}&{}&{}&{}&{}&{}&{}&{}\\[-1.0ex]
$\ell$&$n$,$\kappa<0$,$\kappa>0$&$b_h$&Degenerate states&$E_{n\kappa}(C_{ps}=-5fm^{-1})$&$E_{n\kappa}(C_{ps}=0)$&&&$\ell$&$n$,$\kappa<0$,$\kappa>0$&$b_h$&Degenerate states&$E_{n\kappa}(C_{ps}=5fm^{-1})$&$E_{n\kappa}(C_{ps}=0)$	\\[2.5ex]\hline\hline
1	&1,\ \ \ \ -1,\ \ \ \ 2	&0.005	&$(1s_{1/2},\ \ 0d_{3/2})$	&-6.94839750	&-6.20785454	&&&1	&2,\ \ \ \ -1,\ \ \ \ 2	&0.005	&$(2s_{1/2},\ \ 1d_{3/2})$	&-7.73695190	&-6.79518187	\\[1ex]
	&	&	&	& 1.94858200	& 6.20803901	&&&	&	&	&	& 2.73723990	& 6.79546988	\\[1ex]
	&	&0.010	&	& 1.94578552	&-6.20526456	&&&	&	&0.010	&	& 2.73340961	&-6.79130468	\\[1ex]
	&	&	&	&-6.94504801	& 6.20600224	&&&	&	&	&	&-7.73225609	& 6.79245842	\\[1ex]
	&	&0.015	&	&-6.94138771	& 6.20398859	&&&	&	&0.015	&	& 2.72969057	&-6.78689075	\\[1ex]
	&	&	&	& 1.94305107	&-6.20232449	&&&	&	&	&	&-7.72708905	& 6.78949332	\\[1ex]
	&	&0.020	&	& 1.94037887	& 6.20199793	&&&	&	&0.020	&	&-7.72144673	& 6.78657469	\\[1ex]
	&	&	&	&-6.93741415	&-6.19903082	&&&	&	&	&	& 2.72608325	&-6.78193485	\\[1ex]\hline
2	&1,\ \ \ \ -2,\ \ \ \ 3	&0.005	&$(1p_{3/2},\ \ 0f_{5/2})$	& 3.18693800	& 7.14780707	&&&2	&2,\ \ \ \ -2,\ \ \ \ 3	&0.005	&$(2p_{3/2},\ \ 1f_{5/2})$	& 4.25068600	& 8.02007200	\\[1ex]
	&	&	&	&-8.18681900	&-7.14768804	&&&	&	&	&	&-9.25050700	&-8.01989311	\\[1ex]
	&	&0.010	&	& 3.18293190	&-7.14411991	&&&	&	&0.010	&	& 4.24529380	& 8.01549171	\\[1ex]
	&	&	&	&-8.18245790	& 7.14459395	&&&	&	&	&	&-9.24457950	&-8.01477762	\\[1ex]
	&	&0.015	&	&-8.17790550	&-7.14033441	&&&	&	&0.015	&	& 4.23997740	& 8.01094981	\\[1ex]
	&	&	&	& 3.17897370	& 7.14140281	&&&	&	&	&	&-9.23836760	&-8.00933981	\\[1ex]
	&	&0.020	&	&-8.17316002	&7.138233401	&&&	&	&0.020	&	& 4.23473823	& 8.00644603	\\[1ex]
	&	&	&	& 3.17506373	&-7.13632909	&&&	&	&	&	&-9.23186968	&-8.00357671	\\[1ex]\hline
3	&1,\ \ \ \ -3,\ \ \ \ 4	&0.005	&$(1d_{5/2},\ \ 0g_{7/2})$	&4.341572000	&-8.09663311	&&&3	&2,\ \ \ \ -3,\ \ \ \ 4	&0.005	&$(2d_{5/2},\ \ 1g_{7/2})$	& 5.59430201	&-9.17959900	\\[1ex]
	&	&	&	&-9.34147800	& 8.09672713	&&&	&	&	&	&-10.5941661	& 9.17973513	\\[1ex]
	&	&0.010	&	&-9.33628012	&-8.09217782	&&&	&	&0.010	&	&-10.5871961	&-9.17337521	\\[1ex]
	&	&	&	& 4.33664881	& 8.09254661	&&&	&	&	&	& 5.58773422	& 9.17391330	\\[1ex]
	&	&0.015	&	& 4.33176573	& 8.08838633	&&&	&	&0.015	&	&5.58122603	&-9.16691081	\\[1ex]
	&	&	&	&-9.33093551	&-8.08755590	&&&	&	&	&	&-10.5800132	& 9.16812409	\\[1ex]
	&	&0.020	&	& 4.32692291	&-8.08276641	&&&	&	&0.020	&	&-10.5726151	&-9.16020493	\\[1ex]
	&	&	&	&-9.32544320	& 8.08424650	&&&	&	&	&	& 5.57477690	& 9.16236711	\\[1ex]\hline
4	&1,\ \ \ \ -4,\ \ \ \ 5	&0.005	&$(0f_{7/2},\ \ 0h_{9/2})$	&-10.4561441	& 9.05822302	&&&4	&2,\ \ \ \ -4,\ \ \ \ 5	&0.005	&$(2f_{7/2},\ \ 1h_{9/2})$	&-10.456144	& 9.05822308	\\[1ex]
	&	&	&	& 5.45622503	&-9.05814301	&&&	&	&	&	& 5.45622501	&-9.05814303	\\[1ex]
	&	&0.010	&	& 5.45053720	& 9.05320961	&&&	&	&0.010	&	&5.45053721	& 9.05320961	\\[1ex]
	&	&	&	&-10.4502258	&-9.05289850	&&&	&	&	&	&-10.4502258	&-9.05289850	\\[1ex]
	&	&0.015	&	& 5.44488340	& 9.04821588	&&&	&	&0.015	&	& 5.44488345	& 9.04821511	\\[1ex]
	&	&	&	&-10.4441828	&-9.04751446	&&&	&	&	&	&-10.4441828	&-9.04751442	\\[1ex]
	&	&0.020	&	&-10.4380157	& 9.04323929	&&&	&	&0.020	&	& 5.43926442	& 9.04323923	\\[1ex]
	&	&	&	& 5.43926442	&-9.04199006	&&&	&	&	&	&-10.4380157	&-9.04199015	\\[1ex]\hline\hline
\end{tabular}\label{Tab8}}
\vspace*{-1pt}}
\end{table}
\end{landscape}
\subsubsection{Nonrelativistic limit}
\label{NRL}
Let us now present the non-relativistic limit. This can be achieved when we set $C_s = 0$ and by using the mapping $E_{n\kappa}-M=E_{n_r\ell}$ and $E_{n\kappa}+M=\frac{2\mu}{\hbar^2}$  in equation (\ref{E68}) and (\ref{E69}), then the resulting energy eigenvalues are:
\begin{equation}
E_{n\ell}=\frac{\hbar^2\ell(\ell+1)D_0}{2\mu r_e^2}+D-\frac{\alpha^2\hbar^2}{2\mu r_e^2}\left[\frac{(\delta+n)^2+\frac{\ell(\ell+1)}{\alpha^2c_h^2}(D_1c_h-D_2)+\frac{2\mu Dr_e^2}{\alpha^2\hbar^2}\left(1-\frac{1}{c_h^2}\right)}{2(\delta+n)}\right]^2.
\label{E76}
\end{equation}
It is worth to be paid attention to that despite variation in calculations, equation (\ref{E76}) which is the nonrelativistic limit of the spin Dirac-TW problem and equation (\ref{E45}) which happen to be the nonrelativistic limit of Klein-Gordon-TW problem are exactly the same as the one we obtain for the Schr\"{o}dinger system in equation (\ref{E18}). We have completed the first fold of this work, however, this investigation is not complete without computing information-theoretic quantity.
\section{Information-theoretic measures for TW diatomic molecular potential}\label{sec5}
Recently, the information theory of quantum-mechanical systems have aroused the interest of many Theoretical Physicist \cite{Y7,BJ82, BJ83, BJ84, BJ85, BJ86,BJ87,BJ88,BJ89,BJ90, BJ91, BJ92, BJ93}. This due to the following fact:
\begin{itemize}
\item It provides a deeper insight into the internal structure of the systems \cite{BJ86}.
\item It is the strongest support of the modern quantum computation and information, which is basic for numerous technological developments \cite{BJ87, BJ88}.
\end{itemize}
In this section, the probability distributions which characterize the quantum-mechanical states of TW diatomic molecular potential are analyzed by means of a complementary information measures of a probability distribution called as the Fisher’s information entropy. 

This information measures was originally introduced in 1925 by R. A. Fisher in the theory of statistical estimation \cite{BJ89, BJ90}. This provides the main theoretic tool of the extreme physical information principle and a general variational principle which allows one to derive numerous fundamental equations of physics such as: The Maxwell equations, the Einstein field equations, the Dirac and Klein-Gordon equations, various laws of statistical physics and some laws governing nearly incompressible turbulent fluid flows \cite{Y7,BJ91, BJ92, BJ93}. This probability distribution is defined as
\begin{equation}
I(\rho):=\int\frac{\left[\bar{\nabla}\rho(\bar{{\bf r}})\right]^2}{\rho(\bar{{\bf r}})}d\bar{{\bf r}},
\label{E77}
\end{equation}
where $\bar{\nabla}$ is the gradient operator in polar coordinates. Since equation (\ref{E3}) was obtained from the original Schr\"{o}dinger equation via the variable separable method using the wavefunction $\Psi(r,\theta,\varphi)=G(r)Y_{\ell m}(\theta,\varphi)$ with $G(r)=r^{-1}U_{n\ell}(r)$. We can therefore obtain the radial probability distribution function for this problem as
\begin{eqnarray}
\rho(r)=\aleph e^{-2pb_h(r-r_e)}\left(1-c_he^{-b_h(r-r_e)}\right)^{2q}\left[P_n^{\left(2p,\ \ 2q-1\right)}(1-2c_he^{-b_h(r-r_e)})\right]^2,
\label{E78}
\end{eqnarray}
where $\aleph=N_{n\ell}^2$. The Fisher information entropy of the radial probability distribution function can be calculated by using
\begin{equation}
I(\rho)=\int_0^\infty\frac{1}{\rho(r)}\left[\frac{d\rho({{ r}})}{dr}\right]^2dr=\int_0^{c_he^{b_hr_e}}\frac{1}{\rho(z)}\left[\frac{d\rho(z)}{dz}\right]^2\frac{dz}{b_hz},\ \ \ \ \ \ z=c_he^{-b_h(r-r_e)}.
\label{E79}
\end{equation}
It should be noted that the present calculation becomes rather difficult and too cumbersome due to the interval of the variable $[0, \ c_he^\alpha]$. To overcome this difficulty, as shown in Ref. \cite{Y4,T3}, Wei considered it by weakening the limit of the integral from $[0, \ c_he^\alpha]$ to $[0,\ 1]$ since the solution solved by the modified condition is consistent with the original condition in a high accuracy because of the quickly decreasing factor $(1-z)^q$. If we consider this approximation, we can therefore calculate $I(\rho)$ as
\begin{equation}
I(\rho)=\int_0^\infty\frac{1}{\rho(r)}\left[\frac{d\rho({{ r}})}{dr}\right]^2dr=\frac{1}{2b_h}\int_{-1}^1\frac{1}{\rho(s)}\left[\frac{d\rho({{ s}})}{ds}\right]^2\left(\frac{2}{1-s}\right)ds,\ \ \ s=1-2z.
\label{E80}
\end{equation}
In order to calculate the integral (\ref{E80}), firstly we obtain the first derivative of the radial probability distribution function as
\begin{eqnarray}
&&\frac{d\rho({{ r}})}{dr}=\aleph b_h\left\{2\left(\frac{1-s}{2}\right)^{\tilde{u}}\left(\frac{1+s}{2}\right)^{\tilde{v}+1}\left[P_n^{(\tilde{u},\tilde{v})}(s)\right]^2\left[\frac{q(1-s)}{1+s}-p\right]+\right.\\
&&\left.\frac{1}{2}\left(\frac{1-s}{2}\right)^{\tilde{u}}\left(\frac{1+s}{2}\right)^{\tilde{v}}P_n^{(\tilde{u},\tilde{v})}(s)\left[\frac{n[(\tilde{u}-\tilde{v})-(2n+\tilde{u}+\tilde{v})s]}{2n+\tilde{u}+\tilde{v}}P_n^{(\tilde{u},\tilde{v})}(s)+\frac{2(n+\tilde{u})(n+\tilde{v})}{(2n+\tilde{v}+\tilde{u})}P_{n-1}^{(\tilde{u},\tilde{v})}(s)\right]\right\},\nonumber
\label{E81}
\end{eqnarray}
where we have used the parameters $\tilde{u}=2p$, $\tilde{v}=2q-1$ for simplicity and then utilized the following properties of Jacobi polynomials \cite{BJ64}
\begin{equation}
(1-x^2)\frac{d}{ds}P_n^{(\tilde{u},\tilde{v})}(s)=\left[\frac{n[(\tilde{u}-\tilde{v})-(2n+\tilde{u}+\tilde{v})s]}{2n+\tilde{u}+\tilde{v}}P_n^{(\tilde{u},\tilde{v})}(s)+\frac{2(n+\tilde{u})(n+\tilde{v})}{(2n+\tilde{v}+\tilde{u})}P_{n-1}^{(\tilde{u},\tilde{v})}(s)\right].
\label{E82}
\end{equation}
It is very straightforward calculation to show that
\begin{equation}
\frac{1}{\rho(r)}\left[\frac{d\rho({{r}})}{dr}\right]^2=4b_h^2\aleph \left\{\begin{matrix}\left(\frac{1-s}{2}\right)^{\tilde{u}}\left(\frac{1+s}{2}\right)^{\tilde{v}+1}\left[P_n^{(\tilde{u},\tilde{v})}(s)\right]^2\left[\left(\frac{n(n+\tilde{u})}{2n+\tilde{u}+\tilde{v}}+2q\right)\frac{1}{1+s}-\left(p+q+\frac{n}{2}\right)\right]^2\\
+\left[\frac{(n+\tilde{u})(n+\tilde{v})}{(2n+\tilde{u}+\tilde{v})}\right]\left[\left(\frac{n(n+\tilde{u})}{2n+\tilde{u}+\tilde{v}}+2q\right)\frac{1}{1+s}-\left(p+q+\frac{n}{2}\right)\right]\left(\frac{1-s}{2}\right)^{\tilde{u}}\left(\frac{1+s}{2}\right)^{\tilde{v}}P_n^{(\tilde{u},\tilde{v})}(s)P_{n-1}^{(\tilde{u},\tilde{v})}(s)\\
+\left[\frac{(n+\tilde{u})(n+\tilde{v})}{(2n+\tilde{u}+\tilde{v})}\right]^2\left(\frac{1-s}{2}\right)^{\tilde{u}}\left(\frac{1+s}{\tilde{v}-1}\right)^{\tilde{v}+1}\left[P_{n-1}^{(\tilde{u},\tilde{v})}(s)\right]^2.
\end{matrix}\right.
\label{E83}
\end{equation}
Thus, the Fisher information entropy can be calculated by substituting equation (\ref{E83}) into equation (\ref{E80}) and then decompose into a sum of three integrals to have
\begin{eqnarray}
I(\rho)=2b_h\aleph(I_1+I_2+I_3)
\label{E84}
\end{eqnarray}
with
\begin{eqnarray}
&&I_1= \left[\frac{n(n+\tilde{u})}{2n+\tilde{u}+\tilde{v}}+2q\right]^2I_a+2\left(p+q+\frac{n}{2}\right)\left[\left(p+q+\frac{n}{2}\right)+\left(\frac{n(n+\tilde{u})}{2n+\tilde{u}+\tilde{v}}+2q\right)\right]I_b-\left(p+q+\frac{n}{2}\right)^2I_c\nonumber\\
&&I_2=\left[\frac{(n+\tilde{u})(n+\tilde{v})}{(2n+\tilde{u}+\tilde{v})}\right]^2I_d\ \ \ \mbox{and} \ \ I_3= (n+\tilde{v})\left[\frac{n(n+\tilde{u})}{2n+\tilde{u}+\tilde{v}}+2q\right]I_e-(n+\tilde{v})\left(p+q+\frac{n}{2}\right)I_f\nonumber\\
&&+\left[\frac{n(n+u+\tilde{v})}{2n+\tilde{u}+\tilde{v}}+2q\right]\left(p+q+\frac{n}{2}\right)I_b - \left[\frac{(n+\tilde{u}+\tilde{v})(n+\tilde{v})}{(2n+\tilde{u}+\tilde{v})}\right]\left[\frac{n(n+\tilde{u})}{2n+\tilde{u}+\tilde{v}}+2q\right]I_a,
\nonumber
\end{eqnarray}
where we have utilized the  relation $(n+\tilde{u})P_{n-1}^{\tilde{u},\tilde{v}}(s)+(n+u+v)P_{n}^{\tilde{u},\tilde{v}}(s)-(2n+\tilde{u}+\tilde{v})P_{n}^{\tilde{u},\tilde{v}-1}(s)$ in the calculation of $I_3$ and $I_a, I_b......I_f$ are integrals to be obtain in the next subsection.
\subsection{Evaluation of integrals $I_a$, $I_b$, $I_c$, $I_e$ and $I_f$}
In this section, we calculate the above integrals as follows:
\begin{itemize}
	\item 
	\begin{equation}
	I_a =\int_{-1}^1\left(\frac{1-s}{2}\right)^{\tilde{u}-1}\left(\frac{1+s}{2}\right)^{\tilde{v}-1}\left[P_n^{(\tilde{u},\tilde{v})}(s)\right]^2ds.
	\label{E85}
	\end{equation}
	By using the standard {integral}\footnote{This integral was first derived and used by Ruiz and Dehesa in an attempt to find the Fisher information of orthogonal hypergeometric polynomials \cite{BJ85}} (\ref{A1}), we can find $I_a$ as
	\begin{equation}
	I_a=\frac{n!2^{2n+1}\Gamma(n+2q+2p-1)}{\left[\Gamma(2n+2p+2q)\right]^2}\Gamma(n+2q)\Gamma(n+2p+1)\left(\frac{n+2q-1}{2p}+2+\frac{n+2p}{2q-1}\right).
\label{E86}
\end{equation}
\item
\begin{equation}
	I_b =\int_{-1}^1\left(\frac{1-s}{2}\right)^{\tilde{u}-1}\left(\frac{1+s}{2}\right)^{\tilde{v}}\left[P_n^{(\tilde{u},\tilde{v})}(s)\right]^2ds.
	\label{E87}
	\end{equation}
	By using the standard integral \cite{BJ64} of the appendix (\ref{A2}), $I_b$ can be calculated as
\begin{equation}
	I_b =\frac{\Gamma(2p+n+1)\Gamma(2q+n)}{n!(p+q+n)\Gamma(2p+2q+n)}
	\label{E88}
	\end{equation}
	\item
	\begin{equation}
	I_c =\int_{-1}^1\left(\frac{1-s}{2}\right)^{\tilde{u}}\left(\frac{1+s}{2}\right)^{\tilde{v}}\left[P_n^{(\tilde{u},\tilde{v})}(s)\right]^2ds.
	\label{E89}
	\end{equation}
	Also, by using the standard integral \cite{BJ64} of the appendix (\ref{A2}), $I_b$ can be calculated as
\begin{equation}
	I_c =\frac{\Gamma(2p+n+1)\Gamma(2q+n)}{n!p\Gamma(2p+2q+n)}
	\label{E90}
	\end{equation}
	\item
	\begin{equation}
	I_d =\int_{-1}^1\left(\frac{1-s}{2}\right)^{\tilde{u}-1}\left(\frac{1+s}{2}\right)^{\tilde{v}-1}\left[P_{n-1}^{(\tilde{u},\tilde{v})}(s)\right]^2ds.
	\label{E91}
	\end{equation}
	The above integral can be calculated in a similar fashion as $I_a$. Thus we find
	\begin{equation}
	I_d=\frac{(n-1)!2^{2n-1}\Gamma(n+2q+2p-2)}{\left[\Gamma(2n+2p+2q-2)\right]^2}\Gamma(n+2q-1)\Gamma(n+2p)\left(\frac{n+2q-2}{2p}+2+\frac{n+2p-1}{2q-1}\right)
\label{E92}
\end{equation}
\item
	\begin{equation}
	I_e =\int_{-1}^1\left(\frac{1-s}{2}\right)^{\tilde{u}-1}\left(\frac{1+s}{2}\right)^{\tilde{v}-1}\left[P_{n}^{(\tilde{u},\tilde{v})}(s)\right]\left[P_{n}^{(\tilde{u},\tilde{v}-1)}(s)\right]ds.
	\label{E93}
	\end{equation}
	Since there are no available standard integral in the literature to calculate $I_e$, we therefore resort to splitting the integral into two as
	\begin{equation}
	I_e=(n+v)\left[\frac{n(n+\tilde{u})}{2n+\tilde{u}+\tilde{v}}+2q\right]\left[I_a+\int_{-1}^1\left(\frac{1-s}{2}\right)^{n+\tilde{u}-1}\left(\frac{1+s}{2}\right)^{\tilde{v}-1}\left[P_{n}^{(\tilde{u},\tilde{v})}(s)\right]ds\right],
	\label{E94}
	\end{equation}
	where we have used the following notations of the Jacobi polynomial \cite{BJ64}
	\begin{equation}
	P_{n}^{(\tilde{u},\tilde{v}-1)}(s)=P_{n}^{(\tilde{u},\tilde{v})}(s)+\left(\frac{1-s}{2}\right)^n.
	\label{E95}
	\end{equation}
	Now, by using the integral relation \cite{BJ64} given by appendix (\ref{A4}), we can now calculate $I_e$ as
	\begin{equation}
	I_e=(n+\tilde{v})\left(\frac{n(n+\tilde{u})}{2n+\tilde{u}+\tilde{v}}+2q\right)I_a+I_{\tilde{e}}, 
	\label{E96}
	\end{equation}
	with
	\begin{equation}
	I_{\tilde{e}}= \left[(n+v)\left(\frac{n(n+\tilde{u})}{2n+\tilde{u}+\tilde{v}}+2q\right)\right]\frac{(-1)^n\Gamma(n+2q)\Gamma(2p+n)}{n!q\Gamma(2p+2q-1+n)}\ _3F_2\left(\left.\begin{matrix}-n, 2+2p+2q, 2p+n\\2p+1, 2p+2q+n-1 \end{matrix}\right|1\right).
	\label{E97}
	\end{equation}
	\item
	\begin{eqnarray}
	I_f&=&\int_{-1}^1\left(\frac{1-s}{2}\right)^{\tilde{u}-1}\left(\frac{1+s}{2}\right)^{\tilde{v}}\left[P_{n}^{(\tilde{u},\tilde{v})}(s)\right]\left[P_{n}^{(\tilde{u},\tilde{v}-1)}(s)\right]ds\nonumber\\
	I_f&=&(n+\tilde{v})\left(p+q+\frac{n}{2}\right)\left[I_b+\int_{-1}^1\left(\frac{1-s}{2}\right)^{n+\tilde{u}-1}\left(\frac{1+s}{2}\right)^{\tilde{v}}\left[P_{n}^{(\tilde{u},\tilde{v})}(s)\right]^2ds\right],
	\label{E98}
	\end{eqnarray}
	where we have also used the relation (\ref{E95}). Finally, by using the integral \cite{BJ64} of appendix (\ref{A5})
	\begin{equation}
	I_f=(n+\tilde{v})\left(p+q+\frac{n}{2}\right)I_b+I_{\tilde{f}},
  \label{E99}
	\end{equation}
	with
	\begin{equation}
	I_{\tilde{f}}=2(-1)^n\frac{\Gamma(n+2q)\Gamma(2p+1)\Gamma(n+2p)}{n!\Gamma(n+2p+2q)\Gamma(2p+n+1)\Gamma(1-n)}(n+\tilde{v})\left(p+q+\frac{n}{2}\right).
	\label{E100}
	\end{equation}
  \end{itemize}
\subsection{Main Result}
The Fisher information entropy embedded by the radial probability distribution function can be calculated from equations (\ref{E84})-(\ref{E100}) as
\begin{equation}
I(R)=2b_h\aleph\left\{\begin{matrix}(n+2q)I_a+\left(p+q+\frac{n}{2}\right)\left[\frac{3n(n+2p)}{2n+2p+2q-1}-2+p+9q+\frac{3n}{2}\right]I_b\\
-\left(p+q+\frac{n}{2}\right)^2I_c+\left[\frac{(n+2p)(n+2q-1)}{(2n+2p+2q-1)}\right]I_d+I_{\tilde{e}}+I_{\tilde{f}}.\end{matrix}\right.
\label{E101}
\end{equation}
\section{Numerical Results}\label{sec6}
In table \ref{Tab1}, we present spectroscopic parameters for some diatomic molecules  which are taken from the interesting work of Gordillo-Vizquez and Kunc \cite{T1}. In the case of non-relativistic limit we have calculated the bound-state energy eigenvalues of $H_2 \left(X^1\Sigma^+_g\right)$ and $CO \left(X^1\Sigma^+\right)$ for different values of $n$ and $\ell$ in table (\ref{Tab2}), by using the values in table (\ref{Tab1}). In order to test the accuracy of our results, we compare our obtained energy spectrum with the ones obtained previously in the literature. It is obvious that the present results are in agreement with those obtained by other authors and via other approach. In tables (\ref{Tab3}) and (\ref{Tab4}) we have reported some numerical results for different molecules and states. These molecules are $H_2 \left(X^1\Sigma^+_g\right)$, $CO \left(X^1\Sigma^+\right)$, $HF \left(X^1\Sigma^+\right)$, $O_2 \left(X^3 \Sigma^+_g\right)$, $NO \left(X^2 \Pi_r\right)$ in table (\ref{Tab3}) and  $Cl_2 \left(X^1\Sigma^+_g\right)$, $I_2 \left(X\left(O_g^+\right)\right)$, $N_2 \left(X^1\Sigma^+_g\right)$, $O_2 ^+\left(X^2 \Pi_g\right)$ and $NO^+ \left(X^1 \Sigma^+\right)$ in table (\ref{Tab4}). For the mentioned diatomic molecules, expectation values of $\left\langle V\right\rangle_{n\ell} $, $\left\langle p^2\right\rangle_{n\ell} $, $\left\langle T\right\rangle_{n\ell}  $, $\left\langle T/V\right\rangle_{n\ell} $, $\left\langle r^{-2}\right\rangle_{n\ell}$  are represented in tables (\ref{Tab5}) and (\ref{Tab6}). We have shown the behavior of the energy of non-relativistic limit of TW potential as a function of $b_h$, $\mu$, $c_h$, $r_e$ and $D$ in figures (\ref{fig2}-\ref{fig6}). We have also represented the behavior of energy of TW potential for different values of $n$ and $\ell$ in figures (\ref{fig7}-\ref{fig11}). 

The energy of Dirac equation in the case of spin symmetry limit and pseudospin symmetry limit under the TW diatomic molecular potential is represented in tables (\ref{Tab7}) and (\ref{Tab8}), respectively, for different values of $n$ and $\kappa$. From table (\ref{Tab7}), we can deduce that there are degeneracies between the eigenstates ($np_{1/2}$, $np_{3/2}$), ($nd_{3/2}$, $nd_{5/2}$), ($nf_{5/2}$, $nf_{7/2}$), ($ng_{7/2}$, $ng_{9/2}$), etc. It is worth mentioning that each of these eigenstates form a spin doublet. For example, for any specific value of $n$, where $n = 0, 1, 2,...,$ $np_{1/2}$ with $\kappa = 1$ is considered as the partner of $np_{3/2}$ with $\kappa = −2$. Again, from table (\ref{Tab8}), we can deduce that there are degeneracies between the eigenstates ($ns1/2$, $(n-1)d_{3/2})$, ($np_{3/2}, (n − 1)f_{5/2}$), ($nd_{5/2}, (n-1)g_{7/2}$), ($nf_{7/2}, (n-1)h_{9/2})$, etc. It is worth noting that, each of these eigenstates form a pseudospin doublet. For example, for specific value of $n = 1$, $1s_{1/2}$ with $\kappa = -1$ is considered as the partner of $0d_{3/2}$ with $\kappa = 2$. Thus, states that have pseudo orbital angular momentum $\tilde{\ell}$ quantum numbers, radial $n$ and $n-1$ with $j =\tilde{\ell}-1/2$ and $j =\tilde{\ell}+1/2$, respectively, are degenerate.
	
\section{Conclusions}\label{sec7}
In this paper, we have shown the beauty about eigensolution techniques via an approximate bound state solutions of the Schr\"odinger, Klein-Gordon and Dirac equations for TW diatomic molecular potential. For each type of wave equations and by applying an approximation to the centrifugal term, we obtained the energy eigenvalues and the corresponding wave functions for any quantum state. For further guide to interested readers, we provided some numerical data and figures, which discuss the energy spectrum in each case. The probability distributions of TW diatomic molecular potential were also analyzed. Our obtained solutions find application in various branches of physics and chemistry where non-relativistic and relativistic systems are investigated.

\section*{ Acknowledgements}
The authors wish to thank the referees for their helpful comments and suggestions which have greatly improve the quality of this research work.

\section*{Appendix A: Some useful standard integrals}\label{sec8}
\begin{align}
\int_{-1}^1\left(\frac{1-s}{2}\right)^a\left(\frac{1+x}{2}\right)^b\left[P_n^{(a+1,b+1)}(x)\right]^2dx=\frac{2^{2n+1}n!\Gamma(n+a+2)}{\left[\Gamma(2n+a+b+3)\right]^2}\Gamma(n+b+2)\Gamma(n+a+b+2)\nonumber\\
\times\left(\frac{n+a+1}{b+1}+2+\frac{n+b+1}{a+1}\right).\nonumber\tag{A1}
\label{A1}
\end{align}
\begin{equation}
\int_{-1}^1\left(\frac{1-x}{2}\right)^{a-1}\left(\frac{1+x}{2}\right)^b\left[P_n^{(a,b)}(x)\right]^2dx=\frac{2\Gamma(a+n+1)\Gamma(b+n+1)}{n!a\Gamma(a+b+n+1)}.
\label{A2}
\tag{A2}
\end{equation}
\begin{equation}
\int_{-1}^1\left(\frac{1-x}{2}\right)^{a}\left(\frac{1+x}{2}\right)^b\left[P_n^{(a,b)}(x)\right]^2dx=\frac{2\Gamma(a+n+1)\Gamma(b+n+1)}{n!a\Gamma(a+b+2n+1)\Gamma(a+b+n+1)}.
\label{A3}
\tag{A3}
\end{equation}
\begin{equation}
\int_0^1 x^{\rho-1}(1-x)^{b-\rho-1}\ _2F_1(a,b;c;x)dx=\frac{\Gamma(\rho)\Gamma(\sigma)}{\Gamma(\rho+\sigma)}\ _3F_2(a,b,\rho;\gamma,\rho+\sigma;1).
\label{A4}
\tag{A4}
\end{equation}
\begin{equation}
\int_0^1 x^{\rho-1}(1-x)^{\sigma-1}\ _2F_1(a,b;c;x)dx=\frac{\Gamma(\rho)\Gamma(\gamma)\Gamma(b-\rho)\Gamma(\gamma-a-\rho)}{\Gamma(b)\Gamma(\gamma-a)\Gamma(\gamma-\rho)}.
\label{A5}
\tag{A5}
\end{equation}
\end{changemargin}
\end{document}